\documentclass[aps,showpacs,preprintnumbers,amsmath,amssymb,twocolumn,superscriptaddress,prx,floatfix,nofootbib]{revtex4-2}

\usepackage[export]{adjustbox}
\usepackage{amsfonts} 
\usepackage{amsmath}
\usepackage{amssymb}
\usepackage{graphicx}
\usepackage{dcolumn}
\usepackage{bm}
\usepackage{xkcdcolors}
\usepackage{tabularx}
\usepackage{epstopdf}
\usepackage{xcolor}
\usepackage{mathrsfs}
\usepackage{subcaption}
\usepackage{mathtools}
\usepackage{float}
\usepackage{algpseudocode}
\usepackage{verbatim}
\usepackage{comment}
\usepackage{cleveref}
\usepackage{ragged2e}
\usepackage{multirow}
\usepackage{lineno}

\DeclareCaptionJustification{justified}{\justifying}
\begin{document}

\title{Tracing two decades of carbon emissions using a network approach}

\author{Gianluca Guidi}
\affiliation{IMT School for Advanced Studies, Piazza San Francesco 19, 55100 Lucca (Italy)}
\affiliation{Department of Computer Science, University of Pisa, Largo Bruno Pontecorvo 3, 56127 Pisa (Italy)}
\author{Rossana Mastrandrea}
\email{rossana.mastrandrea@imtlucca.it}
\affiliation{IMT School for Advanced Studies, Piazza San Francesco 19, 55100 Lucca (Italy)}
\author{Angelo Facchini}
\affiliation{IMT School for Advanced Studies, Piazza San Francesco 19, 55100 Lucca (Italy)}
\author{Tiziano Squartini}
\affiliation{IMT School for Advanced Studies, Piazza San Francesco 19, 55100 Lucca (Italy)}
\affiliation{Institute for Advanced Study (IAS), University of Amsterdam, Oude Turfmarkt 145, 1012 GC Amsterdam (The Netherlands)}
\author{Christopher Kennedy}
\affiliation{Institute for Integrated Energy Systems, University of Victoria,\\Victoria, British Columbia (Canada)
}

\date{\today}

\begin{abstract}
Carbon emissions are currently attributed to producers although a consumption-aware accounting is advocated. After constructing the Carbon Trade Network, we trace the flow of emissions over the past two decades. Our analysis reveals the presence of an unexpected, positive feedback: despite individual exchanges have become less carbon-intensive, the increase in trading activity has ultimately risen the amount of emissions directed from `net exporters' towards `net importers'. Adopting a consumption-aware accounting would re-distribute responsibility between the two groups, possibly reducing disparities.
\end{abstract}


\maketitle

\paragraph*{Introduction.} Over the past two decades both the world GDP and total amount of CO2 emissions have increased (see fig. \ref{figA1}). At continent scale, Europe had the largest GDP from 2004 to 2011 when it was surpassed by Asia; in 2015, North America surpassed Europe as well, thus becoming the second-largest world economy. Overall, the Asian GDP has experienced the largest growth throughout the whole time-span (see fig. \ref{figA2}a). Regarding the emissions, the growth rate of Europe and North America has remained close to zero until 2008 and become negative afterwards; Asia, instead, has displayed an increasing trend throughout the entire period (see fig. \ref{figA2}b). Although the two main crises occurred during this period - i.e. the 2008 financial one and the Covid-19 pandemic - had clear consequences on both trends, the observations above suggest that reducing the environmental impact of economic growth remains a challenging goal.

As emissions from burning fossil fuels are the primary cause of global warming~\cite{Canadell2007}, decarbonisation represents the most promising path to undertake for reducing the human impact on the environment: the last decades have witnessed significant efforts to mitigate carbon emissions by, first, identifying the agents to be held accountable for them: at country level, the Intergovernmental Panel on Climate Change (IPPC) accounts emissions according to a `production principle', i.e. by attributing them to countries \emph{producing} goods and services~\cite{IPCC2006,IPCC2019}. 

Recently, however, it has been proposed to consider the adoption of a consumption-aware accounting, prescribing to focus on `final consumers' as well~\cite{Peters2008,Peters2009,Caldeira2010,Caldeira2011}: in other terms, this stream of literature advocates for accounting the carbon embedded into \emph{exported} goods and services, hence re-distributing the responsibility for these emissions between producers and users.

Several authors have attempted to understand to what extent a `consumption principle' may reduce disparities in carbon accounting~\cite{Sun2016,Wu2019}: the estimation of the amount of traded carbon is usually done using Multi-Regional Input-Output (MRIO) tables~\cite{Chen2018,Zhu2018} that, however, are very sensitive to the accuracy of the available data on trading sectors~\cite{Kang2012,Zhuang2019}. As noticed by Caro et al.~\cite{Caro2014}, data requirements can be relaxed by considering aggregate measures: specifically, the amount of carbon embedded into a country export can be quantified by multiplying it by the related carbon intensity of GDP (defined as GDP-CI$=[\text{CO}_2]/\text{GDP}$ and measuring the kilograms of carbon, per dollar of GDP, released by a country during a given year - see Methods and Appendix A): although less accurate, this method overcomes many of the aforementioned problems while maintaining the uncertainty accompanying estimations in a suitable range for quantitative analysis~\cite{Caro2017}.

Inspired by the work of Caro et al., we have adopted a complex network approach~\cite{Barabasi2002,Boccaletti2006,Squartini2018,Cimini2021} and constructed the Carbon Trade Network (CTN), i.e. the graph induced by the trade-embedded carbon exchanges between world countries. Our aim is investigating its complex architecture over the past two decades, to gain insight into the `economic-environmental profile' of each country. Specifically, we will focus on i) the evolution of the GDP-CI of each country and of its nearest neighbours, ii) the geographic distribution of the differences between consumed and produced emissions at country scale, iii) the direction and magnitude of fluxes within and between groups of countries.\\

\begin{figure*}[t!]
\centering
\includegraphics[width=0.49\textwidth,valign=t]{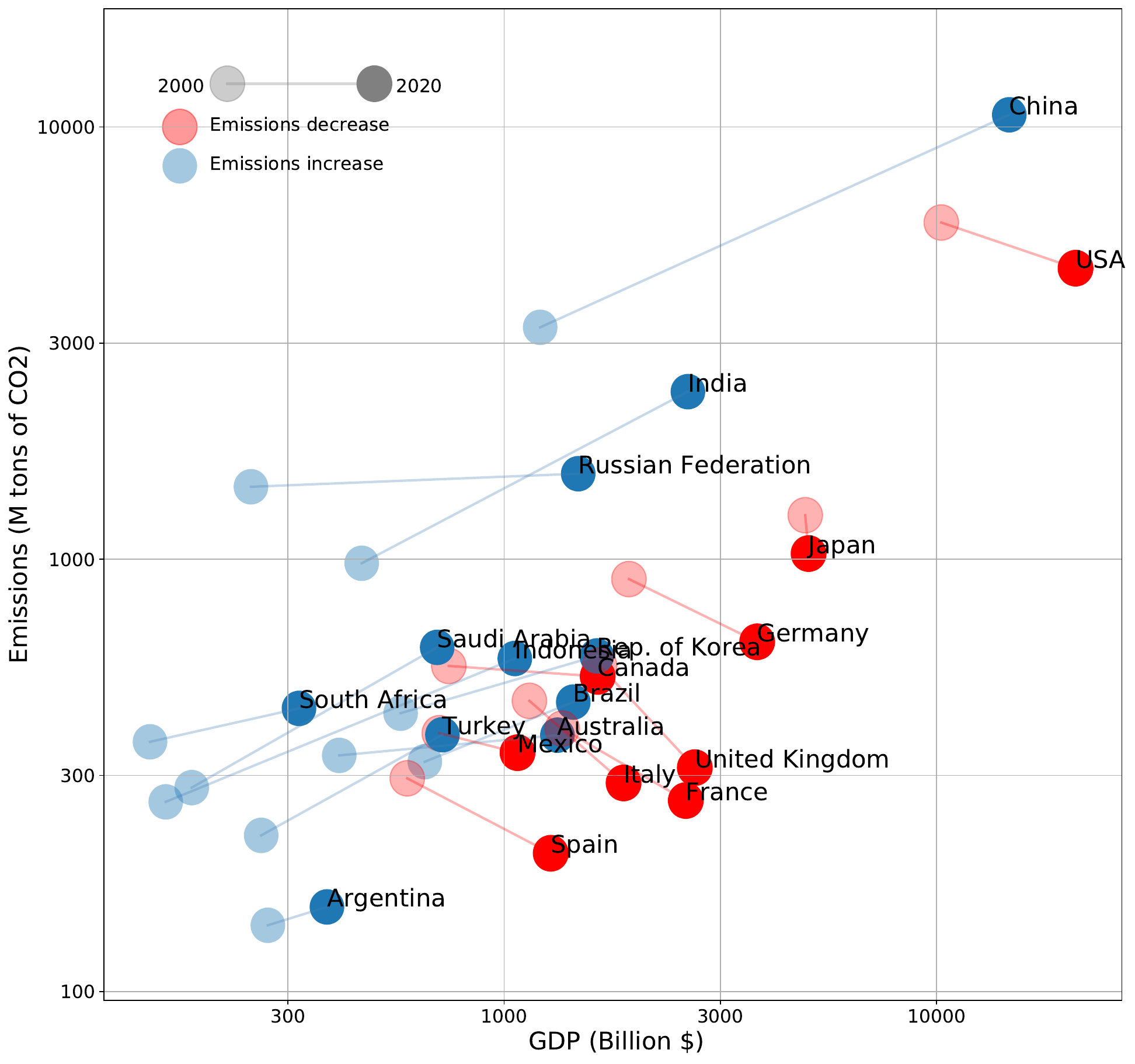}
\includegraphics[width=0.465\textwidth,valign=t]{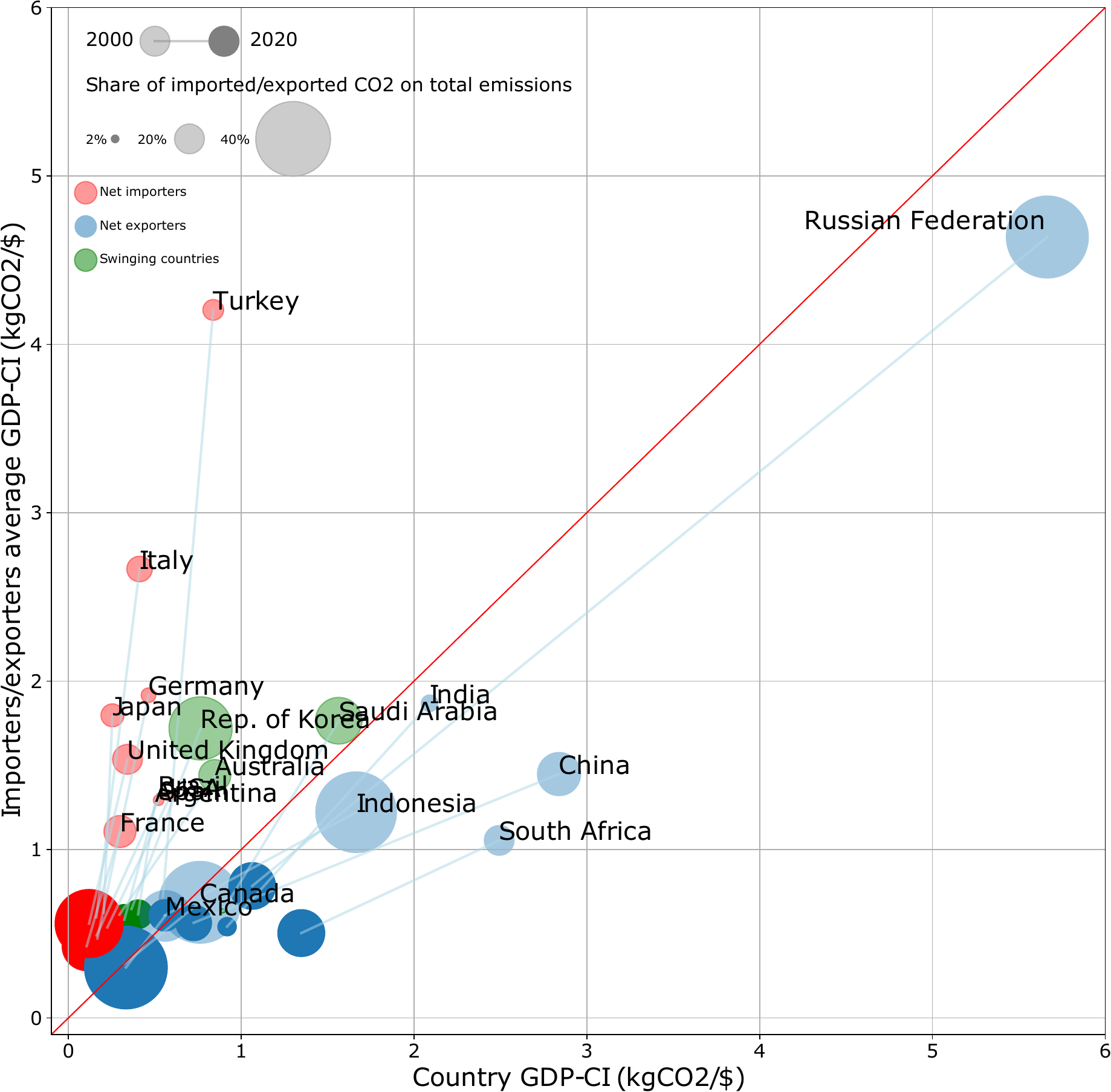}
\caption{Left panel: the `economic-environmental trajectory' of each country emerges upon scattering the tons of carbon released by it versus its GDP, in a yearly fashion. For G20 countries, two tendencies can be identified: the one characterising countries whose GDP and amount of emissions are positively correlated (i.e. Argentina, Brazil, China, India, Indonesia, Korea, the Russian Federation, Saudi Arabia, South Africa and Turkey - moving towards the top-right of the plane) and the one characterising countries whose GDP and amount of emissions are negatively correlated (i.e. France, Germany, Italy, Japan, Spain, United Kingdom and the US - moving towards the bottom-right of the plane). Right panel: scattering the weighted mean of the GDP-CIs of each country exporting partners versus its own GDP-CI reveals that many `net exporters' are less economically efficient than the countries they export to; analogously, many `net importers' are more economically efficient than the countries they import from. Overall, this leads us to conclude that countries whose export exceeds the import, export towards `cleaner' countries; equivalently, countries whose import exceeds the export, import from `less clean' countries. The size of `net exporter' $i$ is proportional to $[t^{out}]_i^y/\text{PE}_i^y$, i.e. the amount of exported emissions over the amount of produced emissions; the size of `net importer' $i$ is proportional to $[t^{in}]_i^y/\text{PE}_i^y$, i.e. the amount of imported emissions over the amount of produced emissions. Numbers are plotted on a doubly logarithmic scale. Names of countries indicate the last year covered by our dataset, i.e. 2020.}
\label{fig1}
\end{figure*}

\paragraph*{Results.} Scattering the tons of carbon released by a nation versus its GDP sheds light on the environmental impact of its economic development. For G20 countries, we observe two, different kinds of evolution (see figs. \ref{fig1}, \ref{figA3}a and \ref{figA3}b): the first one characterises countries whose values of GDP and amount of emissions are positively correlated, i.e. Argentina, Brazil (although their trend shows an inversion in 2018 and 2014, respectively), China, India, Indonesia, the Russian Federation (although the overall amount of its emissions has risen at a much lower rate than others'), Saudi Arabia, South Africa and Turkey; the second one characterises countries whose values of GDP and amount of emissions are negatively correlated, i.e. France, Germany, Italy, Japan, Spain, United Kingdom and the US. Finally, the Australian, Canadian, Korean and Mexican GDPs have risen as well: however, the amount of Canadian emissions has remained quite constant over time; the Australian and Korean ones have risen up to 2008 and 2011, respectively, and become flat afterwards; the Mexican one has risen up to 2012 and decreased afterwards.\\

The analysis carried out so far merely depicts the evolution of aggregate indicators (see also fig. \ref{figA4}). To disentangle the role played by trade, the network representation of the carbon embedded into the exchanges between world countries comes in help. In what follows, we will focus on the CTN, whose generic link weight $c_{ij}$ is defined by multiplying the corresponding trade exchange, say $w_{ij}$, by the GDP-CI of the exporting country: in formulas, $c_{ij}\equiv w_{ij}\cdot\text{GDP-CI}_i$. From a network perspective, the out-strength of node $i$, defined as $t_i^{out}=\sum_{j(\neq i)=1}^Nc_{ij}$, quantifies the total amount of its exported emissions and the in-strength of node $i$, defined as $t_i^{in}=\sum_{j(\neq i)=1}^Nc_{ji}$, quantifies the total amount of its imported emissions; overall, then, the total weight $W=\sum_{i=1}^Nt_i^{out}=\sum_{i=1}^Nt_i^{in}$ quantifies the total amount of trade-embedded carbon emissions (see Methods and Appendix A).

As fig. \ref{figB1} shows, the total weight of the CTN has decreased throughout the second half of our time span, a result indicating that, from 2011 onwards, the impact of trade on world emissions has progressively diminished: one may be, thus, tempted to conclude that the rising trend in fig. \ref{figA1} is solely due to the carbon emitted for internal production; as we will see, this is only partially true.

\begin{figure*}
\centering
\includegraphics[width=\textwidth]{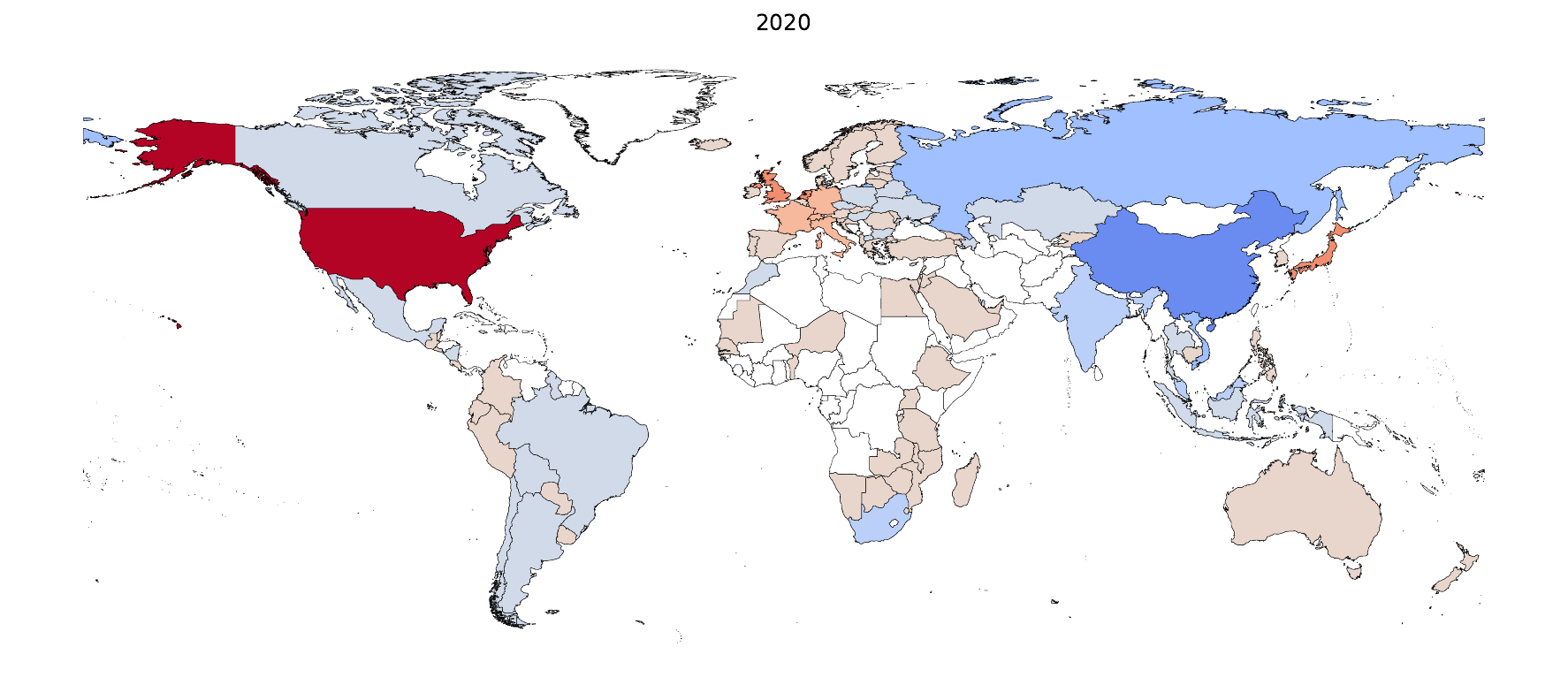}\\
\includegraphics[width=\textwidth]{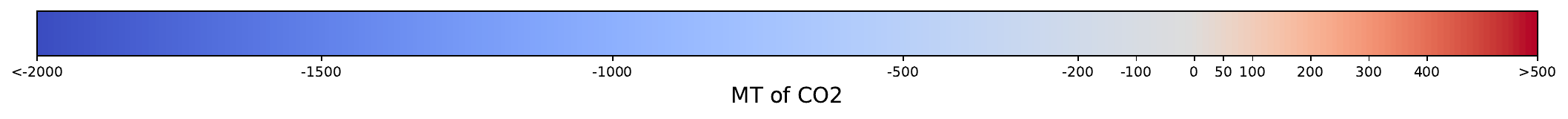}\\
\includegraphics[width=0.49\textwidth,valign=t]{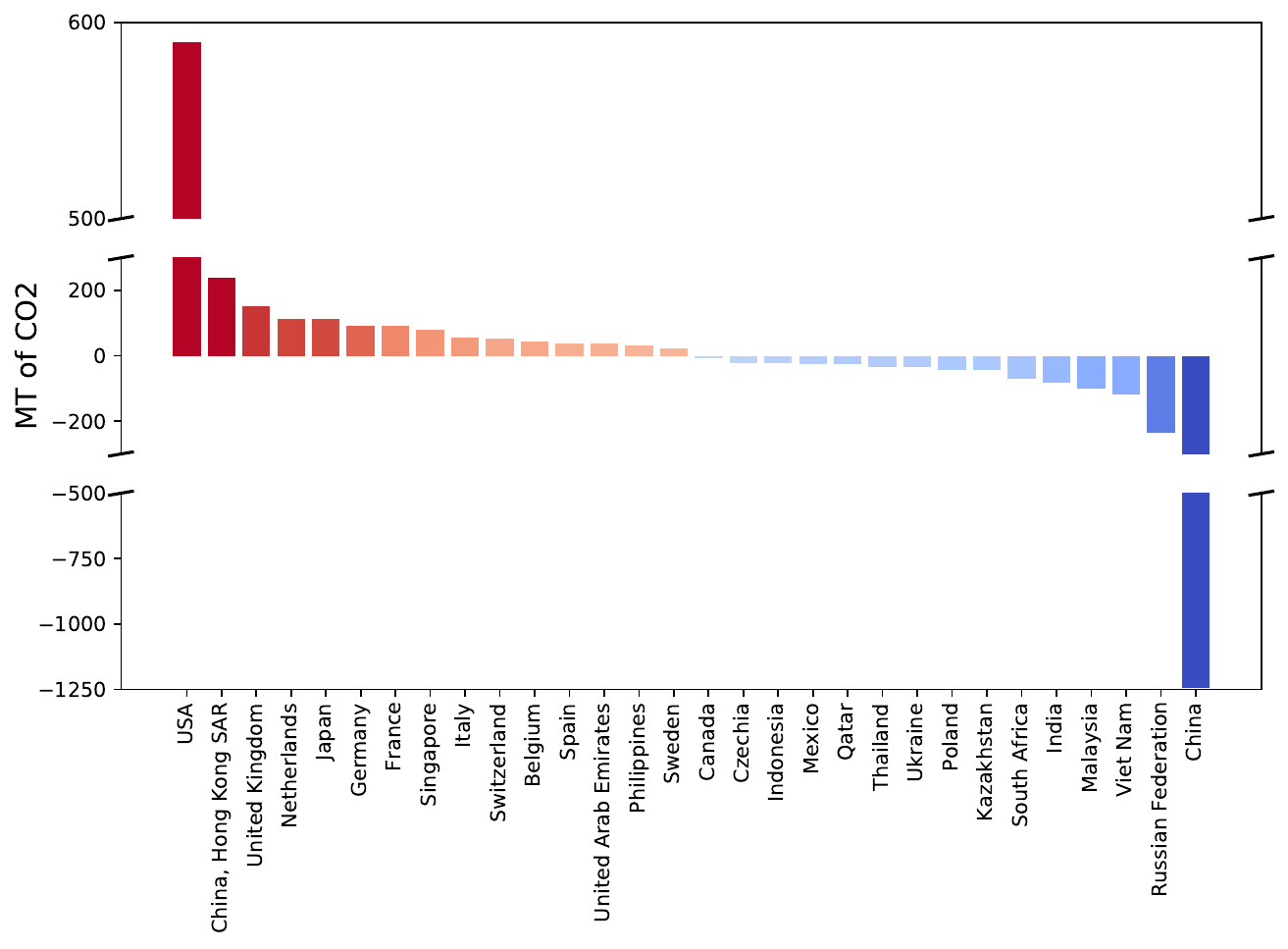}
\includegraphics[width=0.49\textwidth,valign=t]{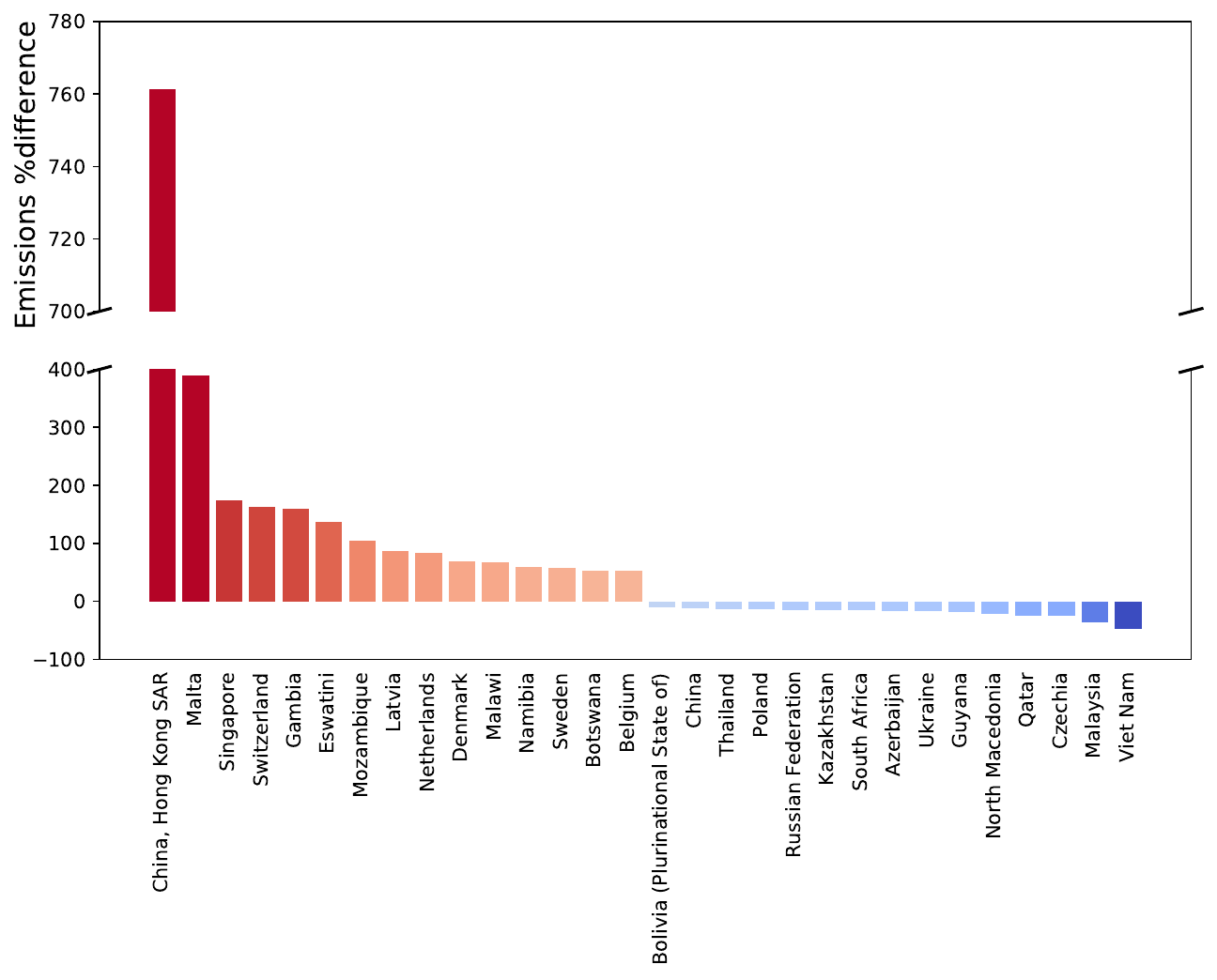}
\caption{Top panel: geographic distribution of the differences between consumed and produced emissions, defined as $\text{CE}_i^{2020}-\text{PE}_i^{2020}\equiv\Delta_i^{2020}$, $\forall\:i$: while `net importers' (or `consumers') of emissions are depicted in shades of red, `net exporters' (or `producers') of emissions are depicted in shades of blue. Bottom-left panel: histogram of the differences between consumed and produced emissions, during the year 2020. Our analysis reveals that France, Germany, Italy, Japan, the UK and the US are among the top ten `net consumers' while China, India, the Russian Federation and South Africa are among the top ten `net producers' - a classification that is robust across time. Bottom-right panel: histogram of the percentage differences between consumed and produced emissions, defined as $[\text{CE}_i^{2020}-\text{PE}_i^{2020}]/\text{PE}_i^{2020}\equiv\Delta_i^{2020}/\text{PE}_i^{2020}$, $\forall\:i$. The ranking changes because of the normalisation. Top `net consumers' are, in this case, countries whose internal production is low (e.g. islands) while top `net producers' are countries with a low level of import.}
\label{fig2}
\end{figure*}

In order to unambiguously classify each country on the basis of its trading behaviour, let us follow Caro et al.~\cite{Caro2014,Caro2017} and re-write the amount of emissions consumed by country $i$, during the year $y$, as $\text{CE}_i^y=\text{PE}_i^y+[t^{in}]_i^y-[t^{out}]_i^y$ (see Appendix B): as a consequence, the difference between the amount of consumed and produced emissions by it reads

\begin{equation}\label{eq1}
\text{CE}_i^y-\text{PE}_i^y=[t^{in}]_i^y-[t^{out}]_i^y\equiv\Delta_i^y,
\end{equation}
a relationship allowing us to distinguish `net importers' of emissions, characterised by $\Delta_i^y>0$ (equivalently, $\text{CE}_i^y>\text{PE}_i^y$), `net exporters' of emissions, characterised by $\Delta_i^y<0$ (equivalently, $\text{CE}_i^y<\text{PE}_i^y$), and countries that have reached `trade-embedded carbon neutrality', characterised by $\Delta_i^y=0$ (equivalently, $\text{CE}_i^y=\text{PE}_i^y$). According to eq. \ref{eq1}, `net importers' (`net exporters') can be also classified as `net consumers' (`net producers') of emissions. Figure \ref{fig2} depicts the distribution of the set of values $\{\Delta_i^{2020}\}$, i.e. of the differences between out-strength and in-strength for each country, during the year 2020: as our analysis reveals, France, Germany, Italy, Japan, the UK and the US are among the top ten `net importers' while China, India, the Russian Federation and South Africa are among the top ten `net exporters'. In order to check the robustness of this classification over time, let us consider the sign of the temporal average $\overline{\Delta_i}\equiv\sum_{y=2000}^{2020}\Delta_i^y/21$, allowing us to distinguish the countries that have mainly served as `net exporters' (i.e. for which $\overline{\Delta_i}<0$) from the countries that have mainly served as `net importers' (i.e. for which $\overline{\Delta_i}>0$): within G20, Argentina, Canada, China, India, Indonesia, Korea, Mexico, the Russian Federation, Saudi Arabia and South Africa belong to the first group while Australia, Brazil, France, Germany, Italy, Japan, Spain, Turkey, the UK and the US belong to the second group (see also fig. \ref{figB2}).\\

Let us, now, study the mutual connections between the two, aforementioned groups of countries. More specifically, let us compare the behaviour of country $i$ (be it a `net exporter' or a `net importer') with that of its partners, by calculating the average value of the GDP-CIs of its neighbours. To this aim, we have distinguished the nodes pointed by it (i.e. the countries it exports to) from the nodes pointing towards it (i.e. the countries it imports from). Figure \ref{fig1} shows the evolution of both kinds of trajectories. Let us focus on `net exporters', first: their trajectories lie below the identity line, a result indicating that their GDP-CI is steadily larger than the one of the countries they export to; for what concerns `net importers', instead, the converse is true: their trajectories lie above the identity line, a result indicating that their GDP-CI is steadily smaller than the one of the countries they import from. Overall, this suggests the presence of a flux of emissions, directed from the countries for which $\overline{\Delta_i}<0$ towards the countries for which $\overline{\Delta_i}>0$ (see also fig. \ref{figB5}).

To gain further insight into this, let us analyse the evolution of the total amount of carbon emissions embedded into the export of `net-exporters'. As fig. \ref{figB3} shows, both the portion directed towards `net importers' and the one directed towards `net exporters' have increased; still, the share of emissions directed towards `net importers' has diminished, a result indicating that what may be called `exp-to-exp' emissions (i.e. the emissions embedded into the trading relationships directed from `net exporters' towards `net exporters') have become increasingly relevant (see also fig. \ref{figB4}).

When considering the evolution of the GDP-CIs, this result may appear paradoxical: each country has, in fact, reduced its own GDP-induced carbon intensity, the major decrease being observable for `net exporters' - specifically, the Russian Federation, South Africa (whose GDP-CI, in 2020, lies slightly above $1\:\text{kg}/\$$), China, India and Saudi Arabia (whose GDP-CI, in 2020, lies between $1\:\text{kg}/\$$ and $0.6\:\text{kg}/\$$): as a  consequence, one would expect the total amount of their trade-embedded carbon emissions to decrease as well. Its rise, only apparently contradictory, is due to an increase of the trading activity involving those countries, causing the related emissions to grow even though each actor has (individually) become more efficient.\\

\paragraph*{Discussion and policy implications.} The seventeen Sustainable Development Goals (SDGs) represent a `call for action', promoted by the United Nations (UN), to be organised along the following guidelines: `\emph{Ending poverty and other deprivations must go together with strategies that improve health and education, reduce inequality and incentivise economic growth - all while tackling climate change and working towards the preservation of our oceans and forests}'~\cite{UN}. As energy emissions from burning fossil fuels are the primary cause of global warming, decarbonisation is one of the most important strategies to reduce the human impact on the environment.

While limiting ourselves to inspect the evolution of the GDP-CIs leads to the conclusion that each country has improved its economic efficiency, a network analysis of trade flows reveals the presence of a flux of emissions, directed from G20 countries serving as `net exporters' towards G20 countries serving as `net importers', whose magnitude has been rising over the past twenty-one years. In other words, our analysis reveals the presence of an unexpected, positive feedback: the rise of trading activity among countries has caused the amount of emissions to rise as well, although exchanges have (individually) become less carbon-intensive.

As the enforced `production-based' accounting criterion solely penalises the countries belonging to the group of `net exporters', our results raise a question: have Europe, North America and the other `net importers' lowered their emissions because of the adoption of `environment-friendly' technologies or are they taking advantage of the current regulation by off-shoring the production of carbon-intensive goods (i.e. moving it to less regulated countries, to be allowed to produce at zero impact and import from elsewhere at a later stage of production)? In the second case, the global carbon footprint would be left unchanged - if not worsened - since offshoring is typically directed towards technologically underdeveloped, hence highly polluting, countries: as signalled by several sources, this practice seems to characterise France~\cite{Dussaux2020}, the United Kingdom~\cite{ER} and the US~\cite{Dai2021}.

On the other hand, a `consumption-aware' criterion to account emissions would burden both the `net exporting' countries as well as those providing the demand for such an export: its adoption may, thus, help reducing economic disparities by incentivising developing countries to transition towards a cleaner industrial production. The enforcement of the Carbon Border Adjustment Mechanism promoted by the European Union~\cite{EU} goes in this direction.\\

\paragraph*{Data.} To construct the CTN, we have combined (i) data on trade flows from UN-COMTRADE (see \texttt{https://comtradeplus.un.org/}), (ii) data on GDPs from the World Bank (see \texttt{https://data.worldbank.org/}), (iii) data on $\text{CO}_2$ emissions from \texttt{https://ourworldindata.org/}. To consistently compare data over the years 2000-2020, we have selected a panel of 111 countries for which trade information was available for the entire period. To the best of our knowledge, the dataset employed to carry out the present study represents a quite unique example in the literature, for sample size, time span and granularity (OCSE reports usually focus on G20 countries over few years).\\

\paragraph*{Methods.} Following~\cite{Caro2017}, carbon emissions are embedded into trade exchanges by considering the yearly values of each country GDP-CI, defined as $\text{GDP-CI}_i^y=[\text{CO}_2]_i^y/\text{GDP}_i^y$. It measures the kilograms of carbon, per dollar of GDP, released by country $i$, during the year $y$, hence conveying information about a country economic efficiency. The GDP-CI allows us to define the weights loading the connections of the CTN, according to the formula $c_{ij}^y=w_{ij}^y\cdot\text{GDP-CI}_i^y$, $\forall\:i\neq j$, with $w_{ij}^y$ indicating the export, in US dollars, from country $i$ to country $j$, during the year $y$. In the present contribution, we have focused on first-order network properties, such as the out- and in-strengths, and second-order network properties, such as the weighted mean of the GDP-CIs of the nodes pointed by/pointing towards node $i$ (see Appendix B).\\

\paragraph*{Acknowledgements.} This work is supported by the European Union - Horizon 2020 Program under the scheme `INFRAIA-01-2018-2019 – Integrating Activities for Advanced Communities', Grant Agreement n. 871042, `SoBigData++: European Integrated Infrastructure for Social Mining and Big Data Analytics' (\href{http://www.sobigdata.eu}{http://www.sobigdata.eu}). This work is also supported by the project `Network analysis of economic and financial resilience', Italian DM n. 289, 25-03-2021 (PRO3 Scuole) CUP D67G22000130001. RM acknowledges support from the `Programma di Attività Integrata' (PAI) project `Prosociality, Cognition and Peer Effects' (Pro.Co.P.E.), funded by IMT School for Advanced Studies Lucca.

\bibliography{References}

\clearpage

\onecolumngrid

\section*{Appendix A.\\Evolution of GDPs, emissions and carbon-intensities over the years 2000-2020}

\setcounter{figure}{0}
\renewcommand{\thefigure}{A\arabic{figure}}

This appendix is devoted to discuss the evolution of 1) the gross domestic product (GDP), 2) the total amount of carbon emissions, 3) the GDP-induced carbon intensity (GDP-CI) and 4) the electric energy-induced carbon intensity (EE-CI) of world countries, over the years 2000-2020. We have considered three, different geographical scales, i.e. world-wise, continent-wise, country-wise. Before proceeding, let us remind that the GDP-CI is defined as $\text{GDP-CI}_i^y=[\text{CO}_2]_i^y/\text{GDP}_i^y$, hence measuring the kilograms of carbon, per dollar of GDP, released by country $i$, during year $y$, and that the EE-CI is defined as $\text{EE-CI}_i^y=[\text{CO}_2]_i^y/\text{kWh}_i^y$, hence measuring the kilograms of carbon, per kilowatt hour, released by country $i$, during year $y$. While the GDP-CI conveys information about a country economic efficiency, the EE-CI conveys information about a country environmental efficiency.\\

\begin{figure*}[t!]
\centering
\includegraphics[width=\textwidth]{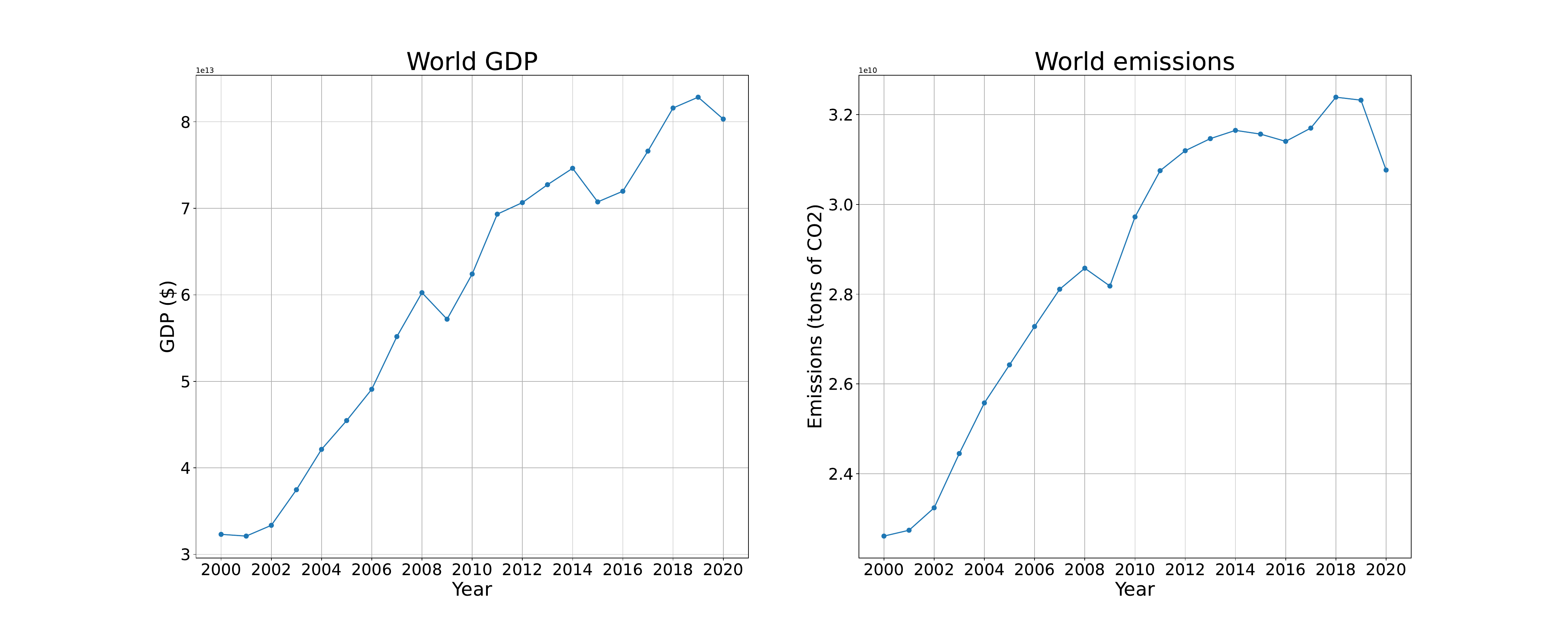}
\caption{Both the world GDP and total amount of $\text{CO}_2$ emissions have risen over the past twenty-one years, suggesting that economic development has led to an increase of the overall amount of carbon released into the atmosphere; still, the impact of the two, main crises occurred within this period of time, i.e. the global financial one (end of 2008) and the Covid-19 pandemic (beginning of 2020), is clearly visible.}
\label{figA1}
\end{figure*}

\paragraph*{World-wise scale.} Let us start by discussing the evolution of the world GDP and total amount of $\text{CO}_2$ emissions. Figure \ref{figA1} shows that both quantities have risen over the past twenty-one years, an evidence suggesting that economic development has led to an increase of the amount of $\text{CO}_2$ released into the atmosphere. Still, the impact of the two main crises occurred within this period of time, i.e. the global financial one (end of 2008) and the Covid-19 pandemic (beginning of 2020), have contributed to slow down the economic growth of countries that, in turn, have reduced their emissions (specifically, in 2009 and 2020).\\

\begin{figure*}[t!]
\centering
\includegraphics[width=\textwidth]{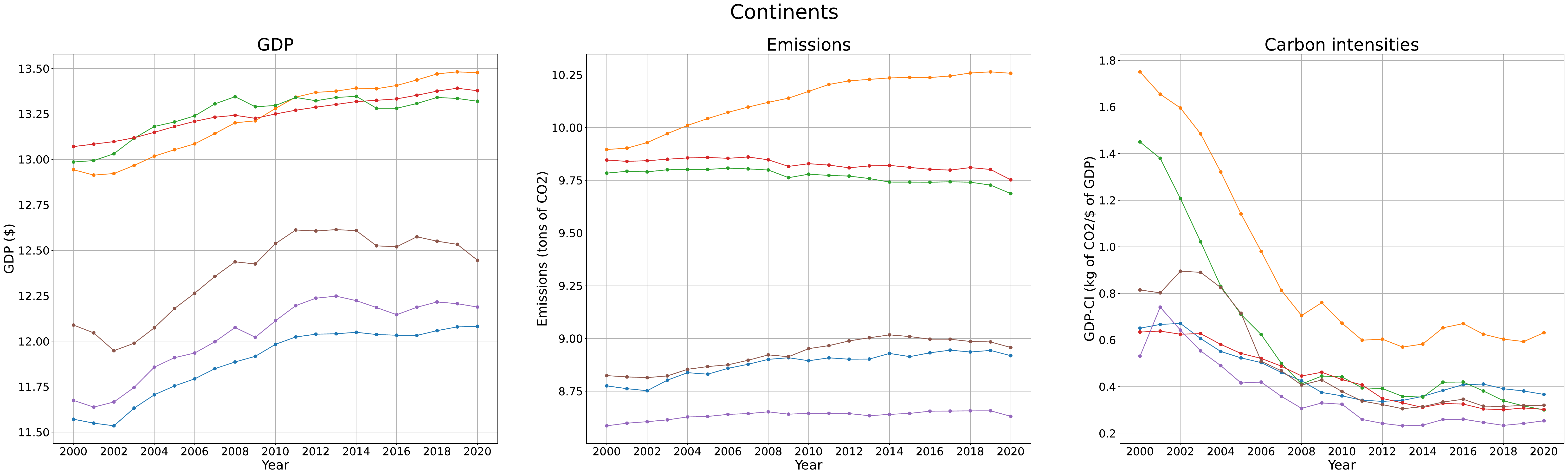}\\
\includegraphics[width=\textwidth]{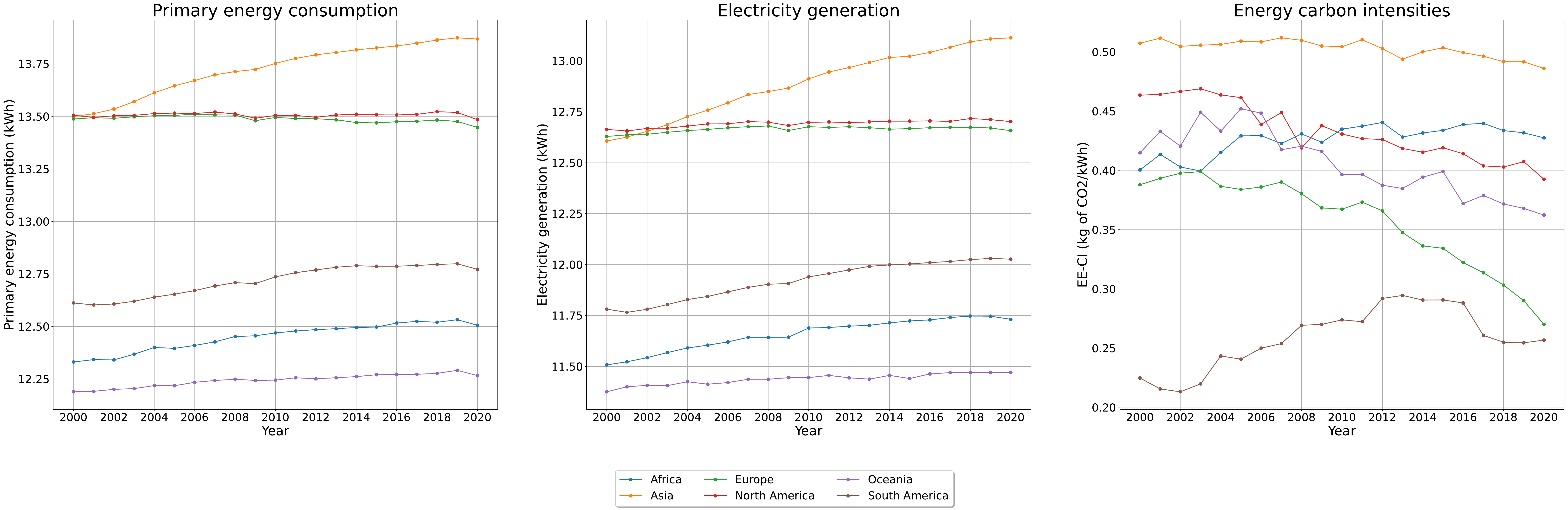}
\caption{All continents have grown during the past twenty-one years. Still, different continents are characterised by different tendencies: Asia, Europe and North America have grown the most, the first one having `surpassed' the other two in 2011. When considering their emissions, however, these three continents show a quite different behaviour: Europe and North America have undertaken a reduction pattern since 2009; Asia, instead, has kept increasing carbon emissions throughout the whole period of time considered here (although at a lower rate since 2011). For what concerns the GDP-CI, each continent has progressively reduced its carbon intensity: however, while Africa, Europe, North America, Oceania and South America have brought their GDP-CI below $0.6\:\text{kg}/\$$ during the years 2004-2006, Asia has kept it above $0.6\:\text{kg}/\$$. For what concerns the EE-CI, instead, Europe, North-America and Oceania have reduced it while Asia has basically kept it constant and Africa and South-America have increased it; conversely, the generation of electric energy has remained constant for all continents except Asia, where it has increased, at a practically constant rate, over the twenty-one years covered by our dataset.}
\label{figA2}
\end{figure*}

\paragraph*{Continent-wise scale.} Let us now `disaggregate' the previous trends in a continent-wise fashion: as fig. \ref{figA2} reveals, continents are roughly split into two groups, i.e. the richest ones (Asia, Europe, North America) and the poorest ones (Africa, Oceania, South America). Overall, the GDP of each continent (calculated as the sum of the GDPs of the countries constituting it) has grown over the past twenty-one years. Interestingly, Europe had the largest GDP from 2004 to 2011 when it stopped growing and was surpassed by Asia, i.e. the continent that has grown the most throughout the whole period of time considered here; North America instead, had the second largest GDP from 2004 to 2009 when it was surpassed by Asia, becoming the third largest one. In 2015, then, the North American GDP surpassed the European GDP becoming, since then, the second largest one.

When coming to consider their emissions, the three richest continents show a quite different behaviour: while Europe and North America have undertaken a reduction pattern since 2009, Asia has kept increasing carbon emissions throughout the whole period of time considered here (although at a lower rate since 2011).

For what concerns the GDP-CI, all continents have reduced their carbon intensity (calculated as the arithmetic mean of the GDP-CIs of the countries constituting it). The Asian GDP-CI, whose value has steadily remained above $0.6\:\text{kg}/\$$, is the largest one. Conversely, Africa, Europe, North America, Oceania and South America have brought their GDP-CI below $0.6\:\text{kg}/\$$ during the triennium 2004-2006. It is worth noticing that while both the Asian GDP and amount of emissions have increased (although at different rates), this is not true for Europe and North America whose GDP has increased while their amount of emissions has decreased.

For what concerns the EE-CI, instead, Europe, North-America and Oceania have reduced it while Asia has basically kept it constant and Africa and South America have increased it. Conversely, the generation of electric energy has remained constant for Europe and North America while all the other continents have increased it - even if with differences - at a practically constant rate, over the twenty-one years covered by our dataset. Taken together, these results indicate that Europe and North-America have produced the same amount of electric energy in a progressively less carbon-intense way while the production of electric energy in all the other continents has increased although its environmental impact has not decreased.\\

\begin{figure*}[t!]
\centering
\includegraphics[width=\textwidth]{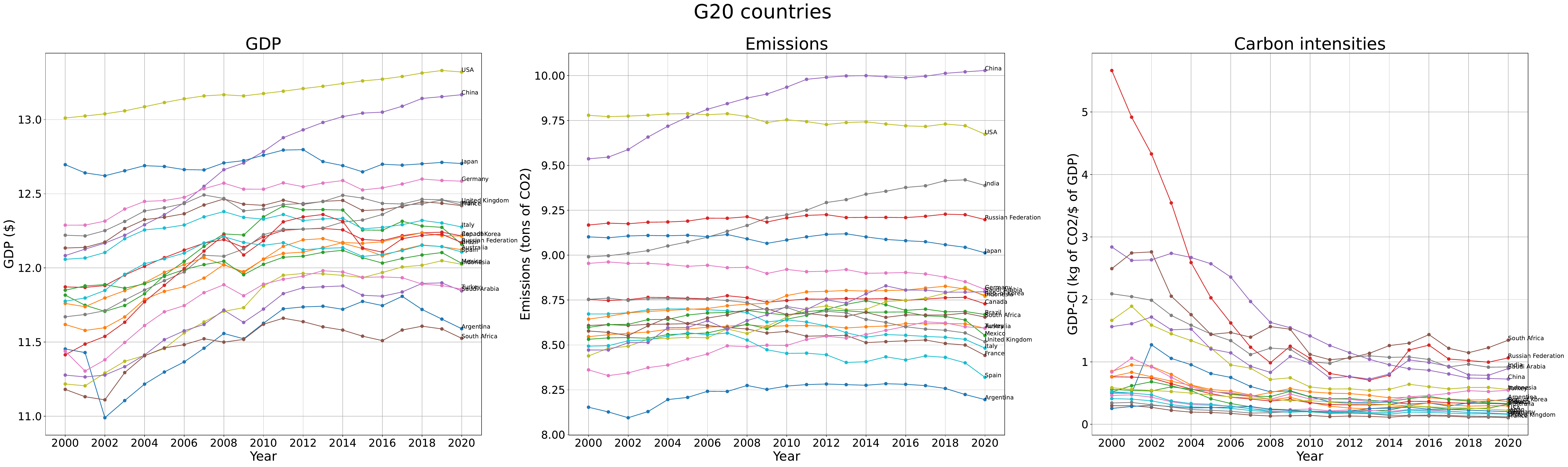}\\
\includegraphics[width=\textwidth]{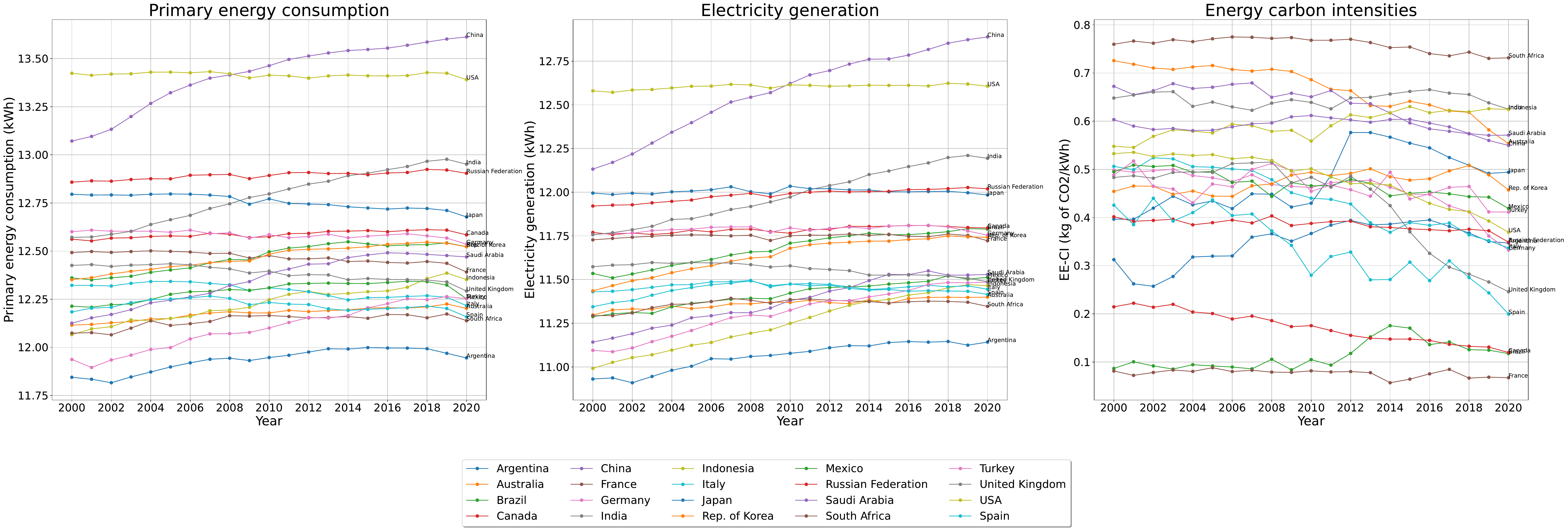}
\caption{All countries have grown during the past twenty-one years. Still, different countries are characterised by different tendencies. The two, far more interesting ones are those characterising China and the US. For what concerns the economic growth, the US remains the leading country; however, while the US emissions have (slowly) started to decrease in 2008, the Chinese ones have kept increasing up to 2012: then, after a stationary trend of almost five years, they have started increasing again. With regards to the GDP-CI, all countries have reduced their carbon intensity: however, the Russian Federation and South Africa have kept it above $1\:\text{kg}/\$$ while China, India and Saudi Arabia have kept it between $1\:\text{kg}/\$$ and $0.6\:\text{kg}/\$$; on the other hand, France, Germany, Italy, Japan, Spain, United Kingdom and the US have kept their GDP-CI below $0.6\:\text{kg}/\$$ throughout the whole period of time considered here. For what concerns the EE-CI, roughly the same groups of countries emerge.}
\label{figA3}
\end{figure*}

\paragraph*{Country-wise scale.} The previous trends can be further `disaggregated' at a country-level. For the sake of illustration, let us focus on G20 countries. Figure \ref{figA3} reveals the co-existence of different tendencies, the two, far more interesting ones being those characterising China and the US. The US remains the leading country in terms of economic growth; however, while the amount of emissions released by the US has slowly started to decrease in 2008, the Chinese one has kept increasing up to 2012: then, after a stationary trend of almost five years, it has started increasing again.

For what concerns the GDP-CI, all countries have reduced their carbon intensity; still, a data-driven threshold clearly emerges suggesting the presence of two, different groups. The first one is composed by China, India, the Russian Federation, Saudi Arabia and South Africa, which have kept their GDP-CI above $0.6\:\text{kg}/\$$ - more precisely, the Russian Federation and South Africa have a GDP-CI which is above $1\:\text{kg}/\$$ while China, India and Saudi Arabia have a GDP-CI lying between $1\:\text{kg}/\$$ and $0.6\:\text{kg}/\$$; the second one is composed by France, Germany, Italy, Japan, Spain, United Kingdom and the US which, instead, have kept their GDP-CI below $0.6\:\text{kg}/\$$ throughout the whole period of time considered here. Indonesia and Turkey `lie in the middle', with Indonesia having lowered its GDP-CI up to the `limiting' value of $0.6\:\text{kg}/\$$ and Turkey having risen it during the last years of our dataset up to the same value. Interestingly, while both the Chinese GDP and amount of emissions have increased, the GDP of the US has increased while its total amount of emissions has decreased.

With respect to the EE-CI, roughly the same groups of countries emerge: the first one is composed by Australia, China, India, Indonesia, Saudi Arabia and South Africa, whose EE-CI has been kept $\gtrsim 0.6\:\text{kg}/\$$ for the vast majority of the temporal snapshots considered here - if not for the entire period; the second one is composed by Canada, France, Germany, Italy, Japan, Korea, Mexico, Spain, the Russian Federation, Turkey, United Kingdom and the US which, instead, have kept their EE-CI below $0.6\:\text{kg}/\$$.

\begin{figure*}[t!]
\centering
\includegraphics[width=\textwidth]{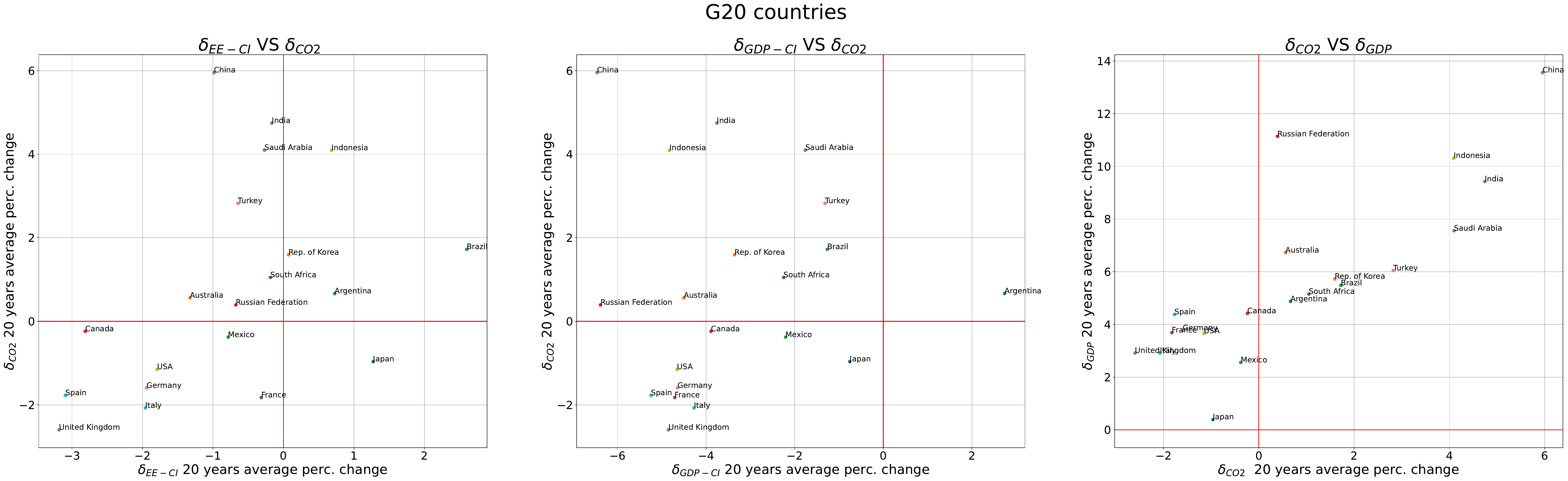}
\caption{Temporal average of the percentual changes of the GDP, the EE-CI and the emissions, for the countries belonging to G20. As our analysis reveals, two sets of countries emerge, i.e. those for which both $\overline{\delta}_\text{GDP}>0$ and $\overline{\delta}_\text{CO2}>0$ (i.e. Argentina, Australia, Brazil, China, India, Indonesia, Korea, the Russian Federation, Saudi Arabia, South Africa, Turkey) and those for which $\overline{\delta}_\text{GDP}>0$ but $\overline{\delta}_\text{CO2}<0$ (i.e. Canada, France, Germany, Italy, Japan, Mexico, Spain, United Kingdom and the US). As our analysis suggests, the countries belonging to the first set export towards the countries belonging to the second set. Moreover, the analysis of the sign of $\overline{\delta}_\text{EE-CI}$ reveals that (only) the countries belonging to the second set have grown in a progressively less carbon-intense way.}
\label{figA4}
\end{figure*}

Figure \ref{figA4} illustrates the results of a more refined analysis, focusing on the temporal average of the percentage changes of the GDP, the EE-CI and the emissions, computed as

\begin{align}
\delta_\text{GDP}^t&\equiv\frac{\text{GDP}_t-\text{GDP}_{t-1}}{\text{GDP}_t},\quad t=2001\dots2020 \label{perc1}\\
\delta_\text{EE-CI}^t&\equiv\frac{\text{EE-CI}_t-\text{EE-CI}_{t-1}}{\text{EE-CI}_t},\quad t=2001\dots2020 \label{perc2}\\
\delta_\text{CO2}^t&\equiv\frac{\text{CO2}_t-\text{CO2}_{t-1}}{\text{CO2}_t},\quad t=2001\dots2020 \label{perc3}
\end{align}
for the countries belonging to G20. Two sets of countries, again, emerge, i.e. those for which both $\overline{\delta}_\text{GDP}>0$ and $\overline{\delta}_\text{CO2}>0$ (i.e. Argentina, Australia, Brazil, China, India, Indonesia, Korea, the Russian Federation, Saudi Arabia, South Africa, Turkey - the developing ones, constituting the BRICS and MIKTA groups) and those for which $\overline{\delta}_\text{GDP}>0$ but $\overline{\delta}_\text{CO2}<0$ (i.e. Canada, France, Germany, Italy, Japan, Mexico, Spain, United Kingdom and the US); notice also that for most of the countries belonging to the first (second) set, $\overline{\delta}_\text{EE-CI}>0$ ($\overline{\delta}_\text{EE-CI}<0$).

To gain further insight into the behaviour of G20 countries, we have divided the area of our subplots in quadrants. Fig. \ref{figA4}c lets four groups of countries emerge: the ones with $\overline{\delta}_\text{CO2}<0$ and $\overline{\delta}_\text{EE-CI}<0$, i.e. Canada, Germany, Italy, Spain, Mexico and the US; the ones with $\overline{\delta}_\text{CO2}<0$ and $\overline{\delta}_\text{EE-CI}>0$, i.e. France and Japan; the ones with $\overline{\delta}_\text{CO2}>0$ and $\overline{\delta}_\text{EE-CI}>0$, i.e. Argentina, Brazil and Indonesia; the ones with $\overline{\delta}_\text{CO2}>0$ and $\overline{\delta}_\text{EE-CI}<0$, i.e. Australia, China, India, Korea, the Russian Federation, Saudi Arabia, South Africa and Turkey. Upon excluding the countries which have employed emissions to rise the production of electric energy - assumed to be one of the main drivers of economic development~\cite{Kennedy2017,Kennedy2018} - we are left with the group of nations which have constantly played the role of `net exporters' (among which the entire group of BRICS with the only exception of Brazil).

Taken together, our results suggest that 1) the countries for which both $\overline{\delta}_\text{GDP}>0$ and $\overline{\delta}_\text{CO2}>0$ export towards the countries for which $\overline{\delta}_\text{GDP}>0$ but $\overline{\delta}_\text{CO2}<0$; 2) (only) the countries belonging to the second set have grown in a progressively less carbon-intense way.

\setcounter{figure}{0}
\renewcommand{\thefigure}{B\arabic{figure}}

\section*{Appendix B.\\Evolution of the Carbon Trade Network over the years 2000-2020}

\begin{figure*}[t!]
\centering
\includegraphics[width=0.32\textwidth]{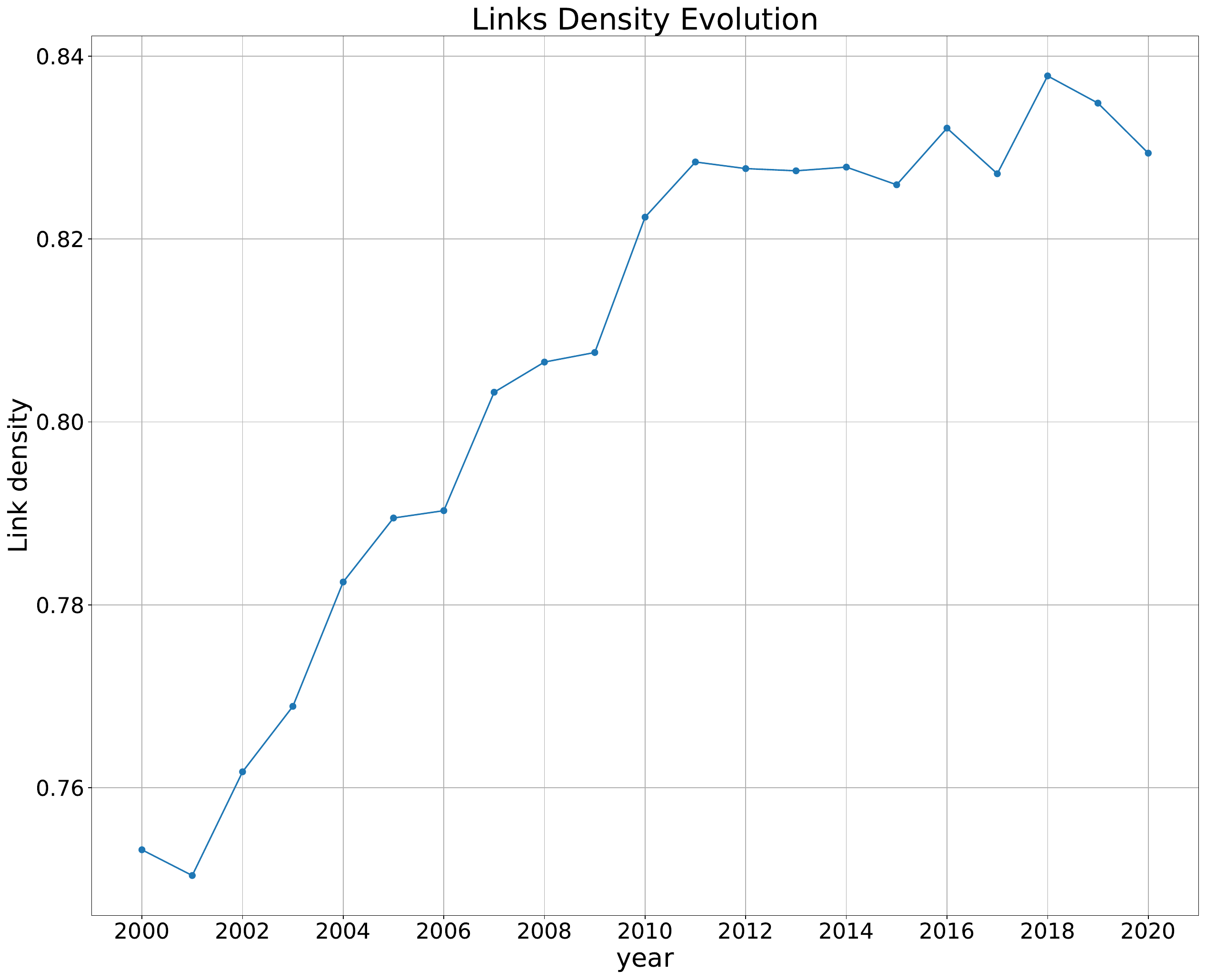}
\includegraphics[width=0.32\textwidth]{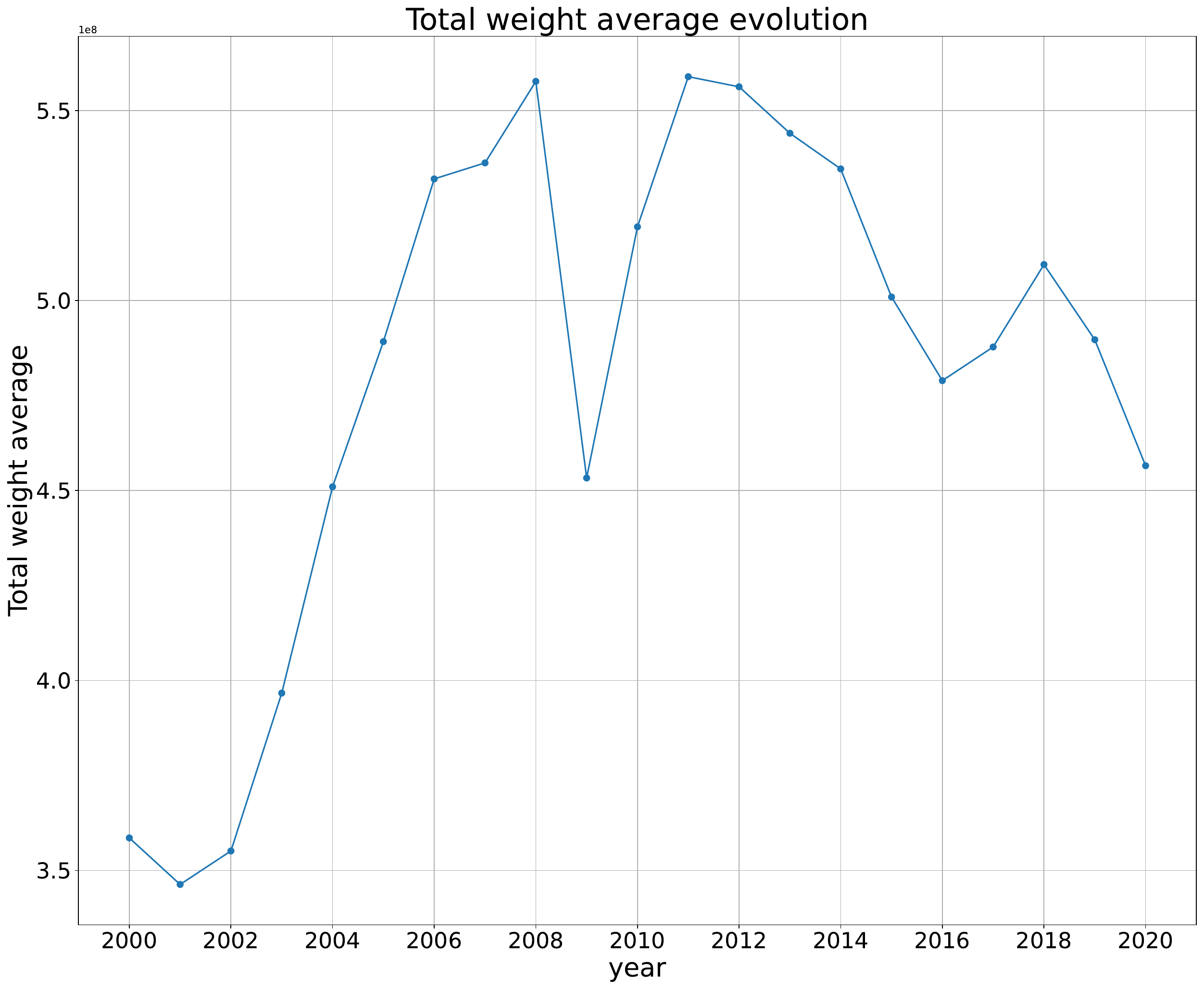}
\includegraphics[width=0.32\textwidth]{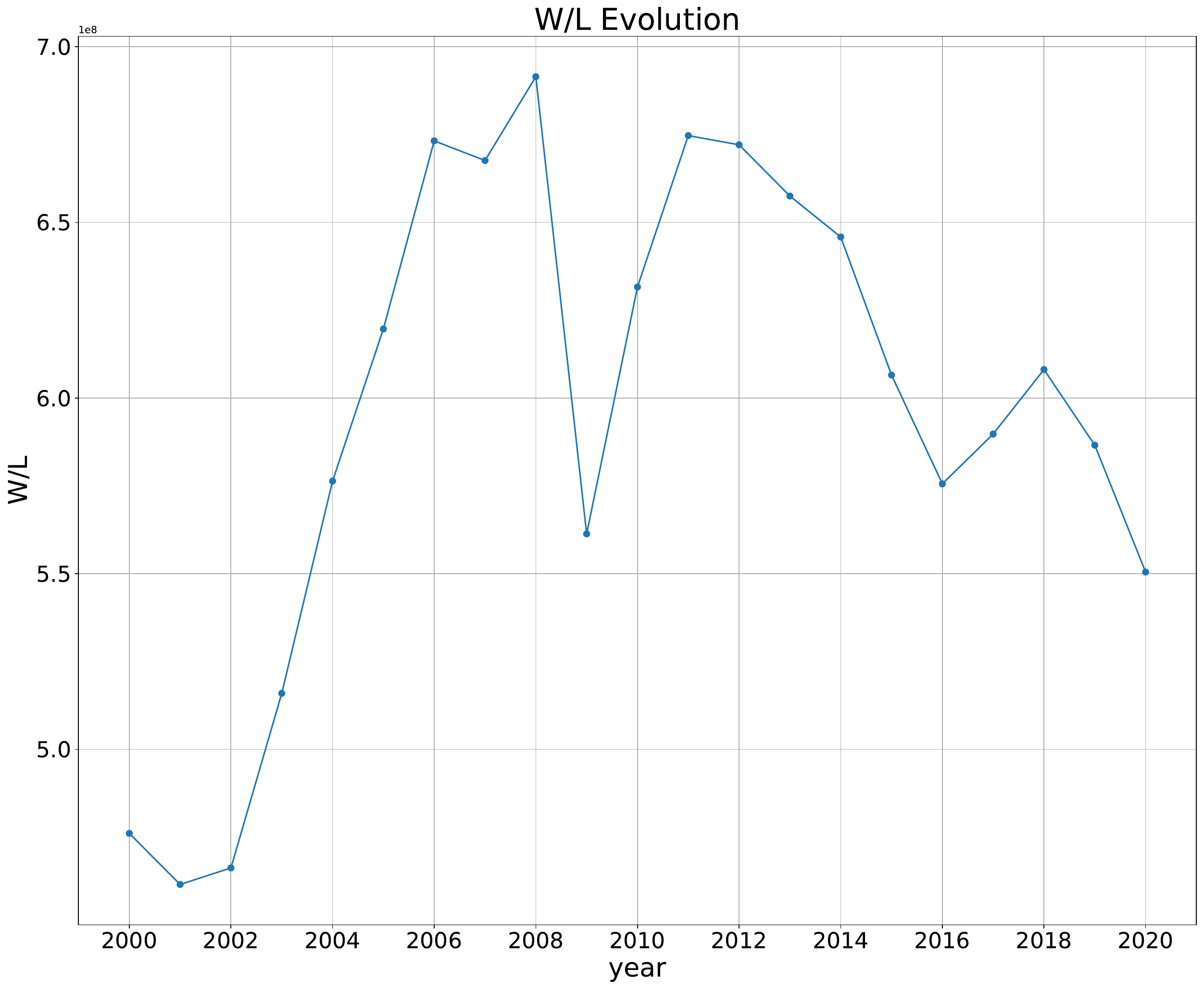}
\caption{Left panel: evolution of the CTN link density, that has risen up to the value of $\simeq0.83$. Middle panel: evolution of the CTN average weight, that has increased from 2002 to 2008; since the network volume is constant, its rise points out that the worldwide amount of emissions embedded into trade exchanges has risen as well. From 2008 on, however, an overall decreasing trend can be appreciated, affected by the two, main crises occurred during this period. Right panel: evolution of the weight-per-link, that confirms that individual exchanges have become less carbon-intensive from 2008 on.}
\label{figB1}
\end{figure*}

This appendix is devoted to discuss the evolution of global network properties over the twenty-one years covered by our dataset. Let us start by defining the link density, reading

\begin{equation}
c=\frac{L}{N(N-1)}
\end{equation}
where $N=111$ and $L=\sum_{i=1}^N\sum_{j(\neq i)=1}^Na_{ij}$ where the generic entry $a_{ij}$ is 1 if $c_{ij}>0$ and zero otherwise. As shown in fig. \ref{figB1}, the CTN link density has risen up to $\simeq0.83$. The weighted counterpart of the link density is the average weight, reading

\begin{equation}
\overline{W}=\frac{W}{N(N-1)}
\end{equation}
where $W=\sum_{i=1}^N\sum_{j(\neq i)=1}^Nc_{ij}$. As shown in fig. \ref{figB1}, the CTN average weight has increased from 2002 to 2008: since the network volume is constant, this means that the total weight, i.e. the worldwide amount of emissions embedded into trade exchanges has risen; since 2008, however, an overall decreasing trend (`affected' by the two, main crises occurred during this period) can be appreciated. Plotting the evolution of the ratio $W/L$ confirms that individual exchanges have become less carbon-intensive.\\

\begin{figure*}[t!]
\centering
\includegraphics[width=\textwidth]{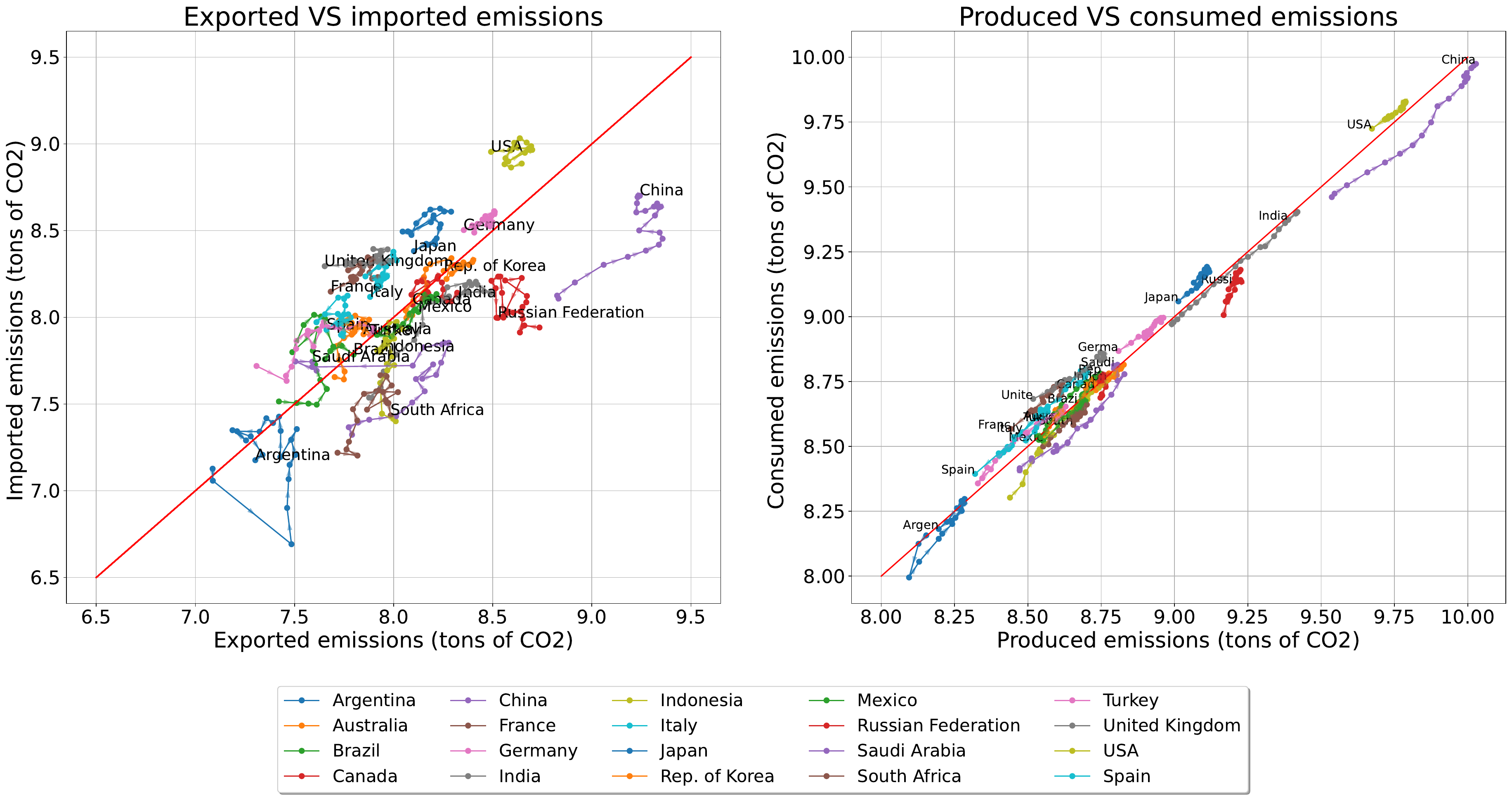}
\caption{Left panel: scattering a country out-strength versus its in-strength reveals its behaviour over the twenty-one years considered here. Argentina, Canada, China, India, Indonesia, Korea, Mexico, the Russian Federation, Saudi Arabia and South Africa are countries whose amount of exported emissions is (overall) larger than the amount of imported emissions; on the other hand, Australia, Brazil, France, Germany, Italy, Japan, Spain, Turkey, the UK and the US are countries whose amount of imported emissions is (overall) larger than the amount of exported emissions. Right panel: the trajectory of G20 countries in the plane defined by scattering the amount of consumed emissions versus the amount of produced emissions. Names of countries indicate the last year covered by our dataset, i.e. 2020.}
\label{figB2}
\end{figure*}

Different accounting criteria impact on different sets of countries: in fact, a context where attributions are production-based disfavours those whose amount of exported emissions is larger than the amount of imported emissions. This consideration suggests us to (try to) classify each country either as a `net producer' or as a `net consumer' of emissions. To this aim, let us consider that the amount of consumed emissions can be re-written as

\begin{equation}
\text{Consumed Emissions}_i^y=\text{Produced Emissions}_i^y+[t^{in}]_i^y-[t^{out}]_i^y
\end{equation}
i.e. as the amount of produced emissions, plus those due to import and minus those due to export - to be attributed to other countries as a consequence of the trading activity `stimulated' by them. From the equation above, it follows that

\begin{equation}
\text{Consumed Emissions}_i^y-\text{Produced Emissions}_i^y=[t^{in}]_i^y-[t^{out}]_i^y
\end{equation}
a relationship establishing that the difference between the total amount of consumed emissions and the total amount of produced emissions by country $i$ matches the difference between its total import and its total export. If

\begin{equation}
\text{Consumed Emissions}_i^y-\text{Produced Emissions}_i^y=[t^{in}]_i^y-[t^{out}]_i^y\geq0
\end{equation}
then, country $i$ is a `net importer', or `net consumer', of emissions; if, on the other hand,

\begin{equation}
\text{Consumed Emissions}_i^y-\text{Produced Emissions}_i^y=[t^{in}]_i^y-[t^{out}]_i^y\leq0
\end{equation}
then, country $i$ is a `net exporter', or `net producer', of emissions. Let us, now, spot the trading behaviour of each country, by scattering its out-strength versus its in-strength: as fig. \ref{figB2} shows, China, India, Indonesia, Mexico, the Russian Federation and South Africa are countries whose amount of exported emissions has steadily exceeded the amount of imported emissions; on the other hand, Brazil, France, Germany, Italy, Japan, Spain, Turkey, the UK and the US are countries whose amount of imported emissions has steadily exceeded the amount of exported emissions. The other G20 countries are `swinging states' whose role changes during the years: specifically, Argentina, Canada, Korea and Saudi Arabia have mainly served as `net exporters' while Australia has mainly served as `net importer'. Notice that the set of `net exporters' roughly matches the group of nations whose GDP-CI is $\gtrsim 0.6\:\text{kg}/\$$.\\

\begin{figure*}[t!]
\centering
\includegraphics[width=0.49\textwidth]{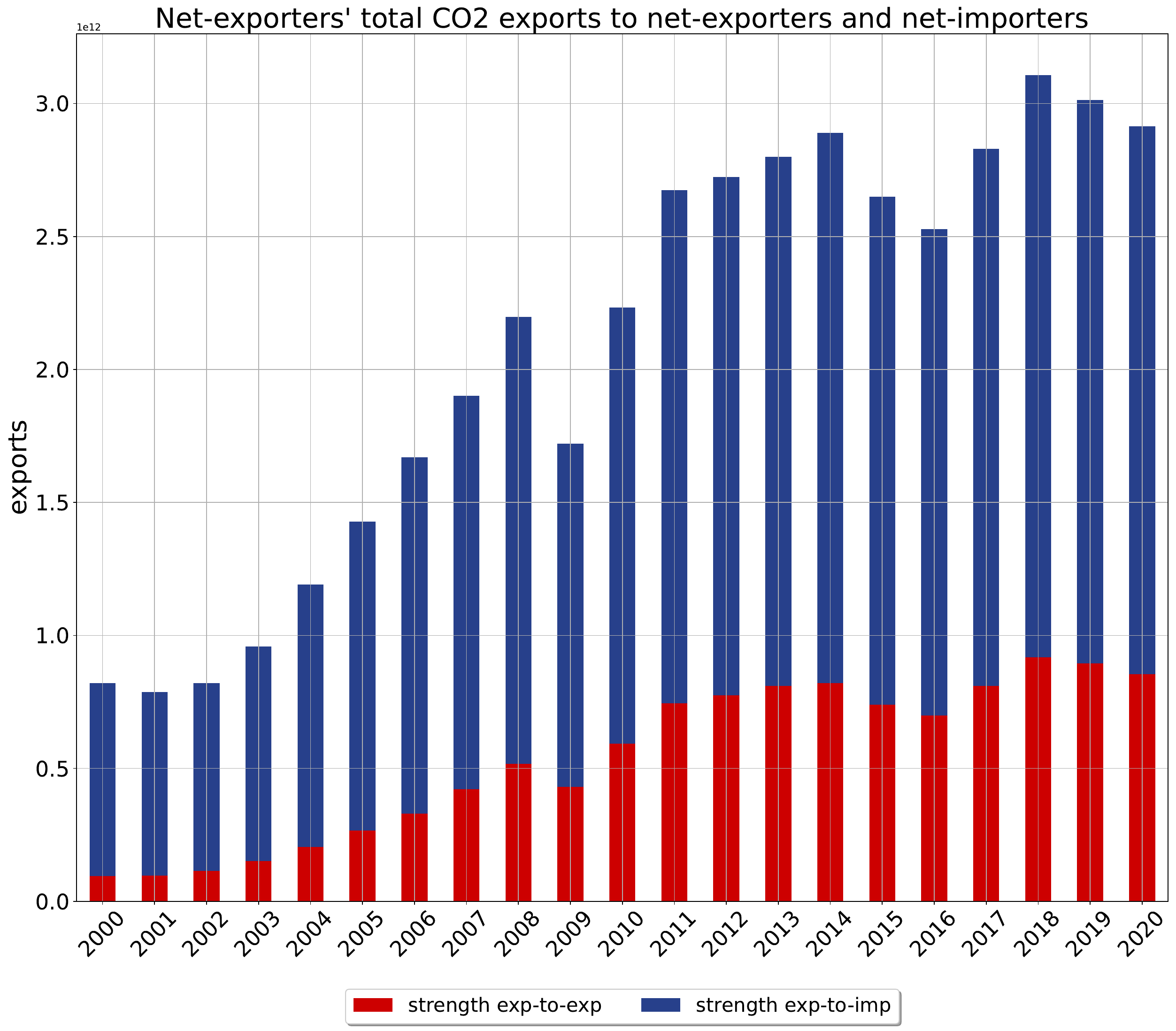}
\includegraphics[width=0.49\textwidth]{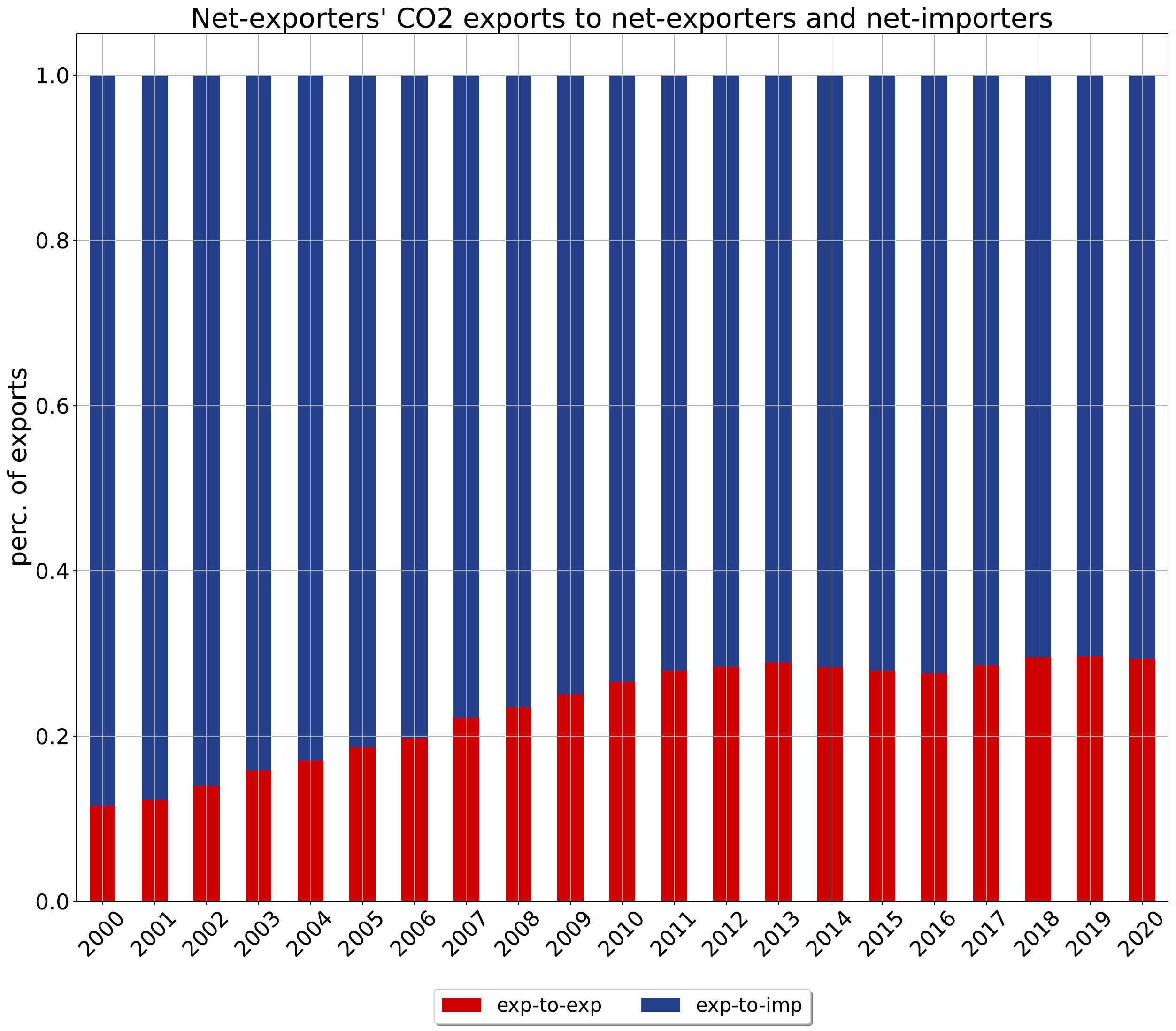}\\
\includegraphics[width=0.49\textwidth]{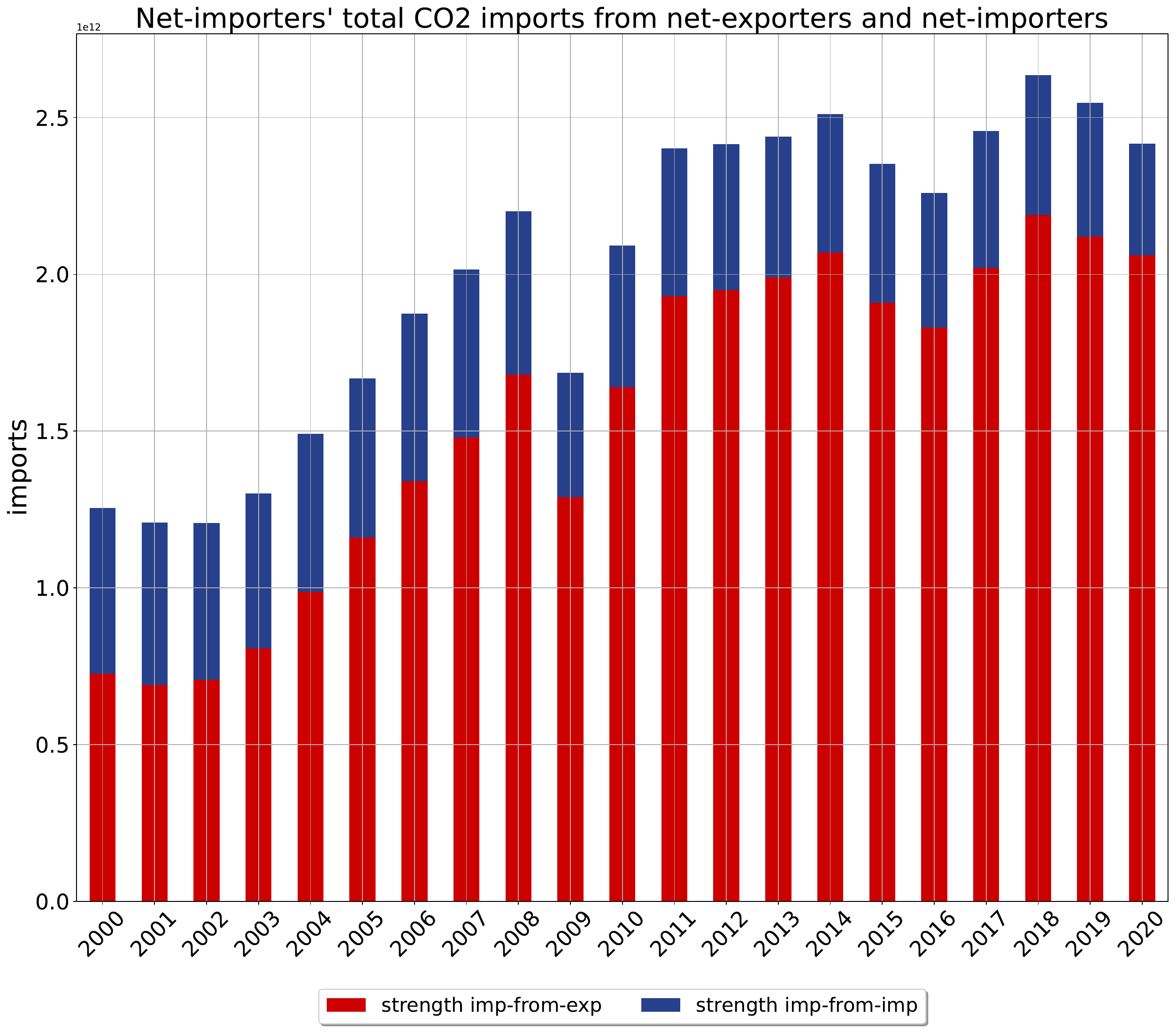}
\includegraphics[width=0.49\textwidth]{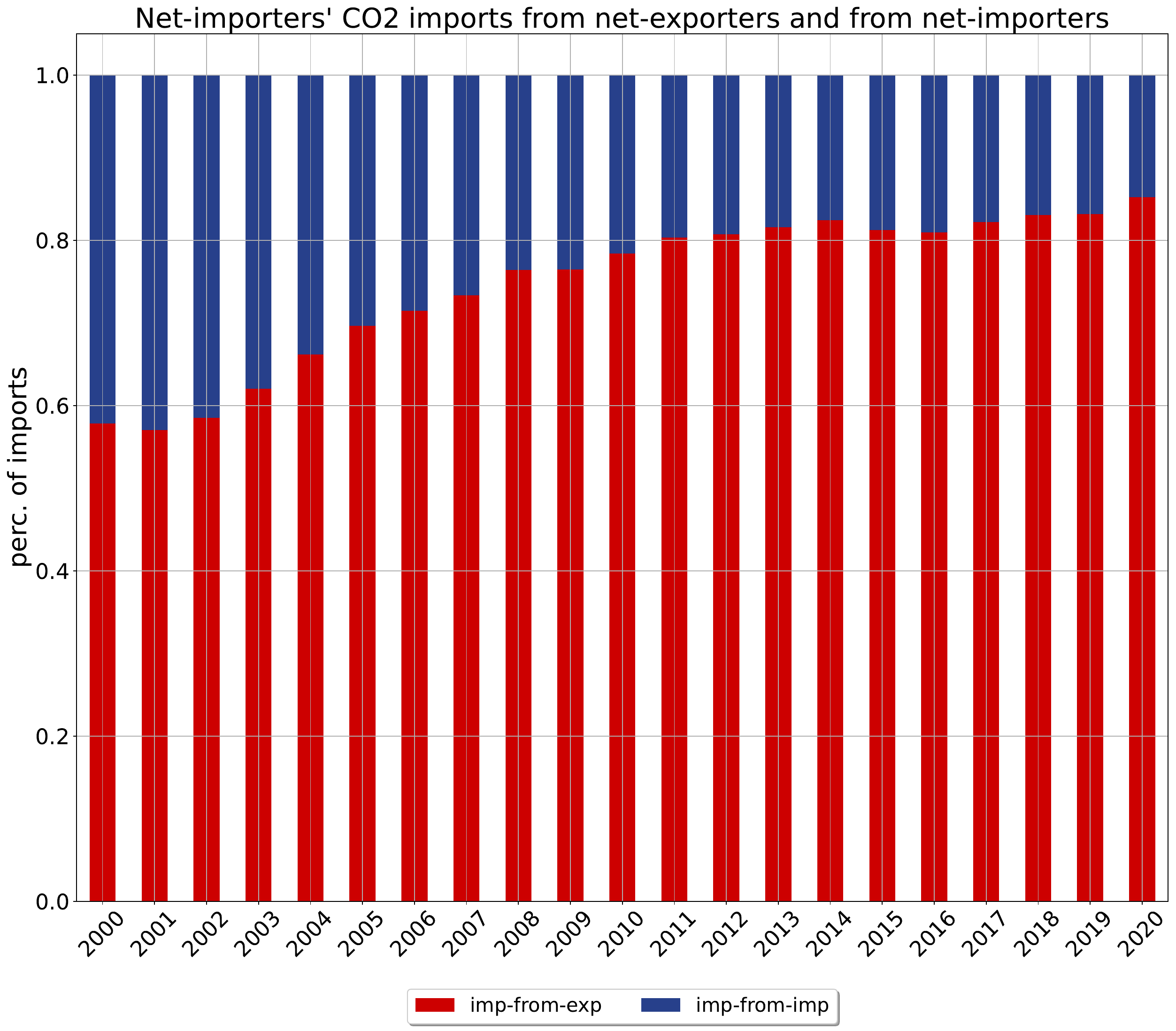}
\caption{Partitioning the group of G20 countries into `net exporters' and `net importers' reveals the presence of a net flux of emissions, directed from the first set of countries towards the second one, whose magnitude has risen over the past twenty-one years. Still, the amount of emissions directed from `net exporters' towards `net exporters' themselves has become increasingly relevant.}
\label{figB3}
\end{figure*}

Partitioning the group of G20 countries into the two, aforementioned sets and studying the evolution of the weight of the connections directed from `net exporters' towards `net importers' leads to fig. \ref{figB3}: as it shows, the total amount of carbon emissions embedded into the export of the first group of countries has, overall, increased. While this is true in absolute terms, the share of emissions directed towards the group of `net importers' has diminished, a result indicating that the role played by the emissions embedded into the trading relationships directed from `net exporters' towards `net exporters' has become increasingly relevant.

\begin{figure*}[t!]
\centering
\includegraphics[width=\textwidth]{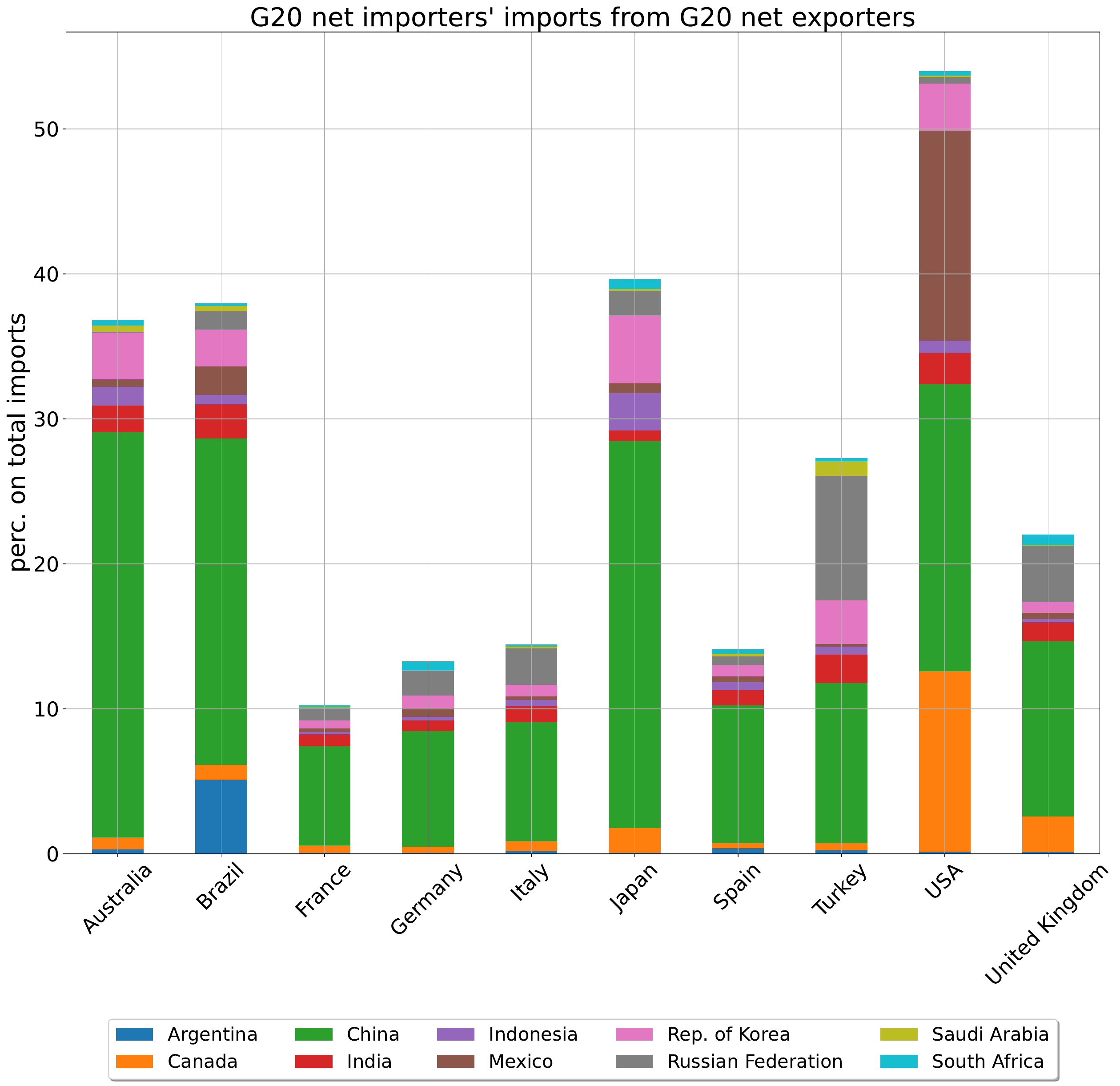}
\caption{Composition of the import basket of each `net consumer' (belonging to G20), partitioned according to the share of emissions coming from each `net producer' (belonging to G20). The role played by `net exporters' of emissions is, in some cases, substantial as it amounts at (more than) the $50\%$ of the total import of the US, $\simeq40\%$ of the total import of Australia, Brazil and Japan, $\simeq30\%$ of the total import of Turkey, $\simeq20\%$ of the total import of the United Kingdom.}
\label{figB4}
\end{figure*}

For the sake of completeness, we have partitioned the import of each `net consumer' according to the share of emissions due to each `net producer': as fig. \ref{figB4} reveals, the role played by the `net exporters' of emissions is, in some cases, substantial, amounting at (more than) the $50\%$ of the total import of the US, $\simeq40\%$ of the total import of Australia, Brazil and Japan, $\simeq30\%$ of the total import of Turkey, $\simeq20\%$ of the total import of the United Kingdom.\\

As we have pointed out in the main text, observing a decrease of the carbon intensities does not necessarily lead to a conclusive answer about the `cleanness' of an agent production (be it a continent or a country) as it may adopt strategies such as delocalisation: the likelihood of such a possibility can be explicitly inspected upon analysing the behaviour of a country neighbours. The results obtained so far suggest us to check the identity of the countries importing from the `net exporters' of emissions: should the former ones have smaller GDP-CI and EE-CI, one may indeed suspect them to take advantage of the current regulation about emissions accounting. To this aim, let us consider network properties capturing the behaviour of a node neighbours: in particular, we can define the weighted average nearest neighbours GDP-CIs as

\begin{align}
\overline{\text{GDP-CI}}^{out}_i&=\frac{\sum_{j=1}^Nw_{ij}\cdot\text{GDP-CI}_j}{\sum_{j=1}^Nw_{ij}}=\frac{\sum_{j=1}^Nw_{ij}\cdot\text{GDP-CI}_j}{s_i^{out}},\quad\forall\:i,\\
\overline{\text{GDP-CI}}^{in}_i&=\frac{\sum_{j=1}^Nw_{ji}\cdot\text{GDP-CI}_j}{\sum_{j=1}^Nw_{ji}}=\frac{\sum_{j=1}^Nw_{ji}\cdot\text{GDP-CI}_j}{s_i^{in}},\quad\forall\:i,
\end{align}
i.e. distinguishing the weighted mean of the GDP-CIs of the nodes pointed by node $i$ from the weighted mean of the GDP-CIs of the nodes pointing towards node $i$. Analogously, we can define the weighted average nearest neighbours EE-CIs as

\begin{align}
\overline{\text{EE-CI}}^{out}_i&=\frac{\sum_{j=1}^Nw_{ij}\cdot\text{EE-CI}_j}{\sum_{j=1}^Nw_{ij}}=\frac{\sum_{j=1}^Nw_{ij}\cdot\text{EE-CI}_j}{s_i^{out}},\quad\forall\:i,\\
\overline{\text{EE-CI}}^{in}_i&=\frac{\sum_{j=1}^Nw_{ji}\cdot\text{EE-CI}_j}{\sum_{j=1}^Nw_{ji}}=\frac{\sum_{j=1}^Nw_{ji}\cdot\text{EE-CI}_j}{s_i^{in}},\quad\forall\:i;
\end{align}
as evident from our definitions, $s_i^{out}=\sum_{j=1}^Nw_{ij}$ is nothing but the purely trade-induced out-strength of node $i$ while $s_i^{in}=\sum_{j=1}^Nw_{ji}$ is nothing but the purely trade-induced in-strength of node $i$.

\begin{figure*}[t!]
\centering
\includegraphics[width=0.49\textwidth]{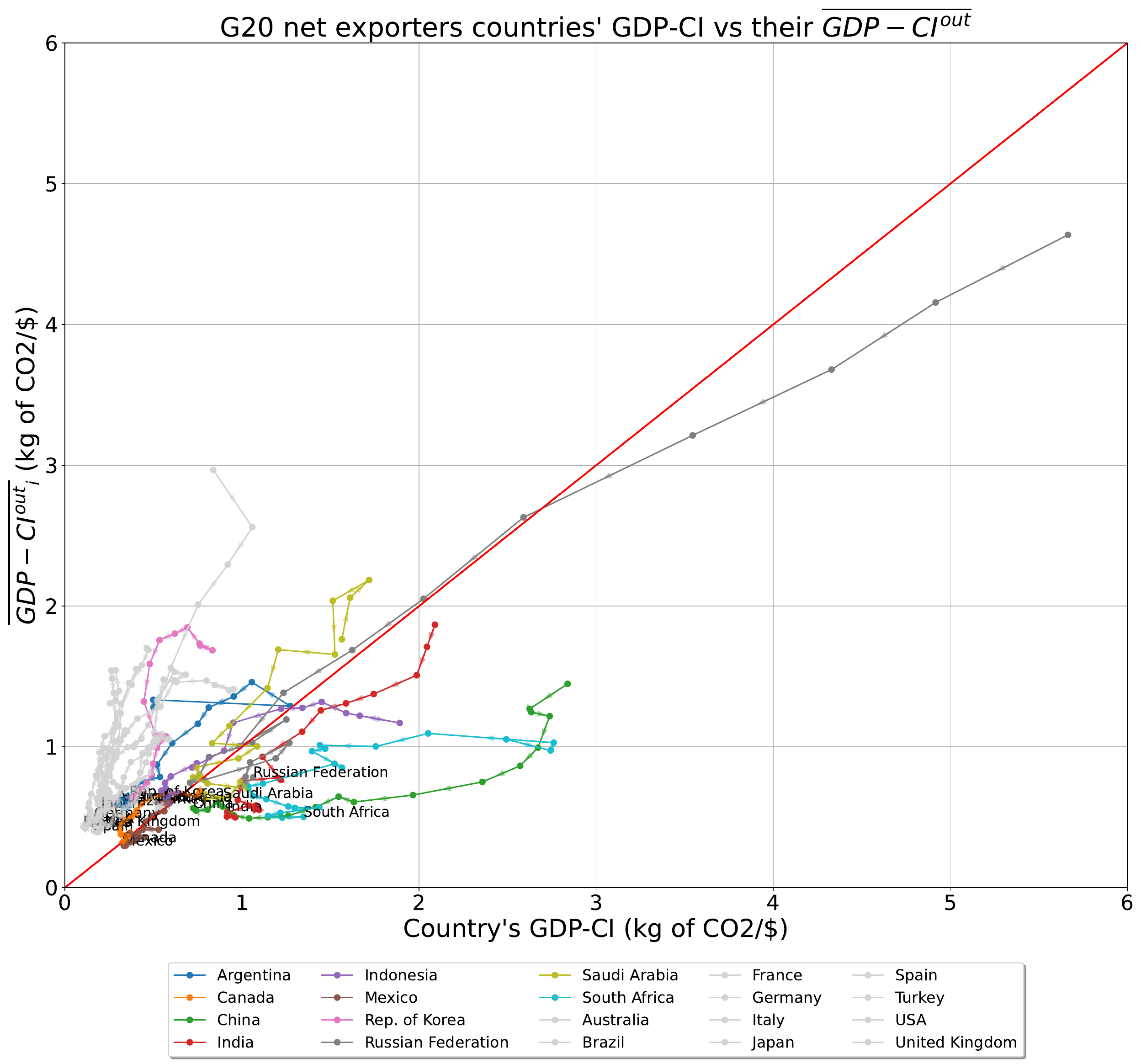}
\includegraphics[width=0.49\textwidth]{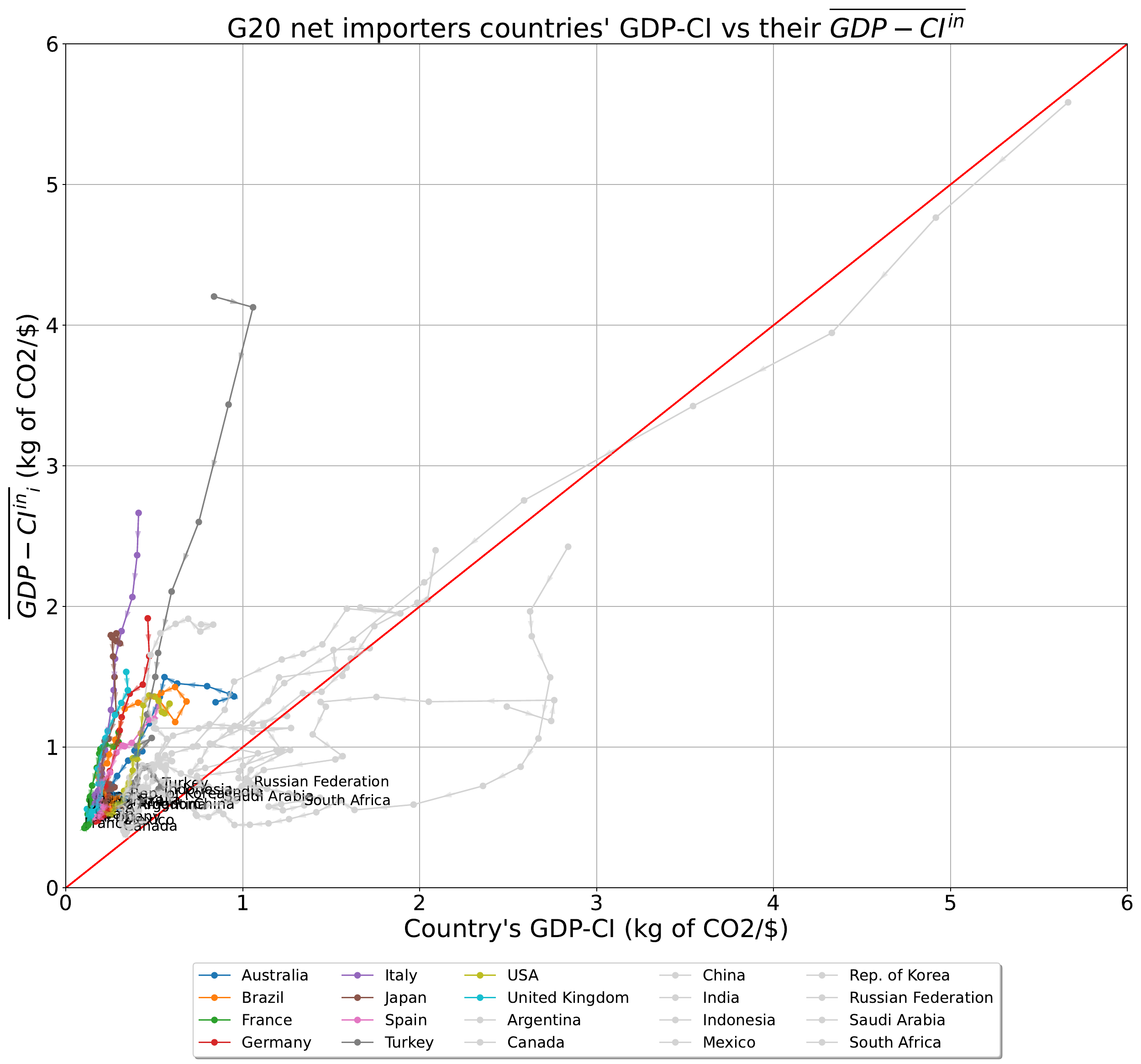}\\
\includegraphics[width=0.49\textwidth]{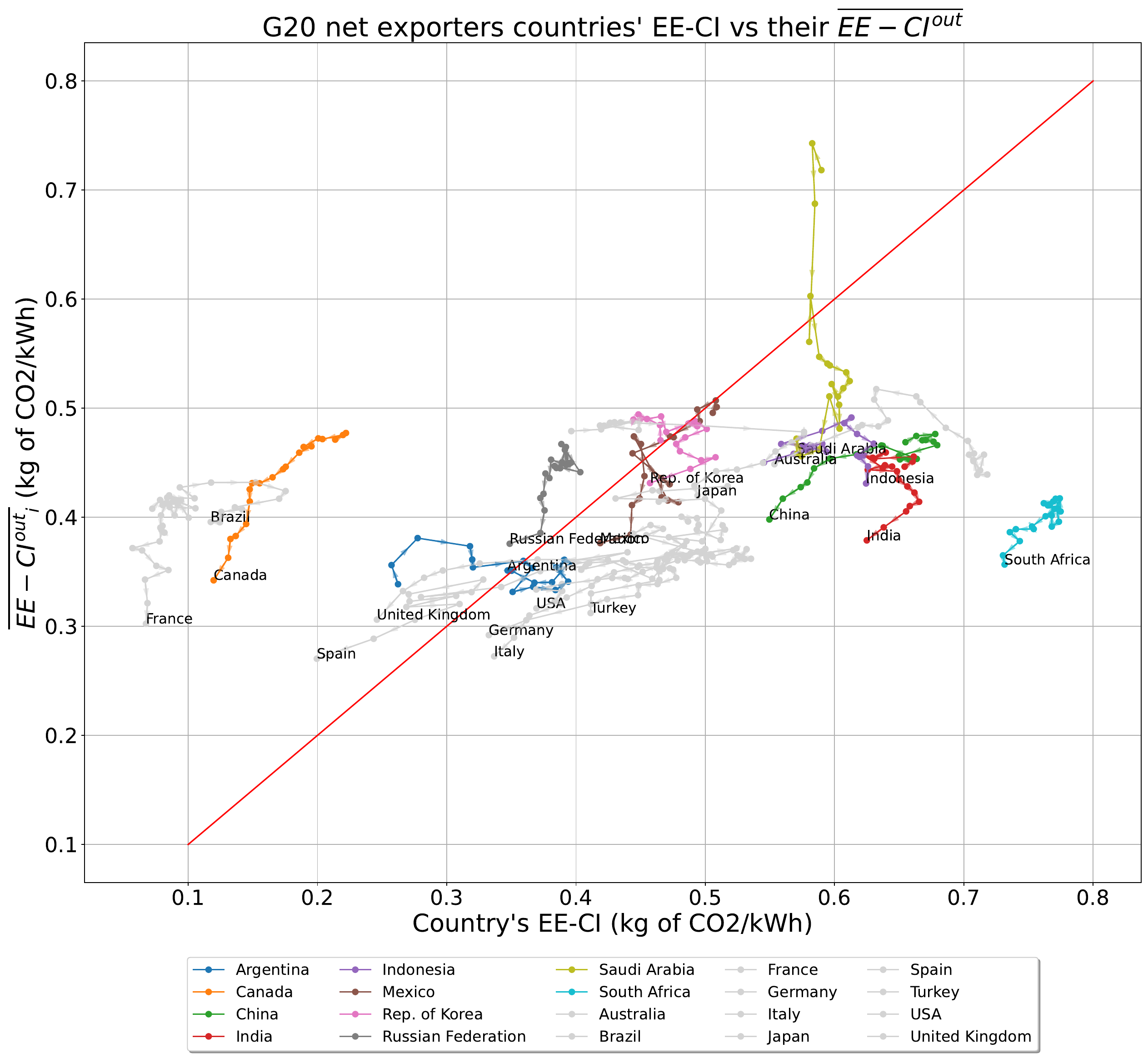}
\includegraphics[width=0.49\textwidth]{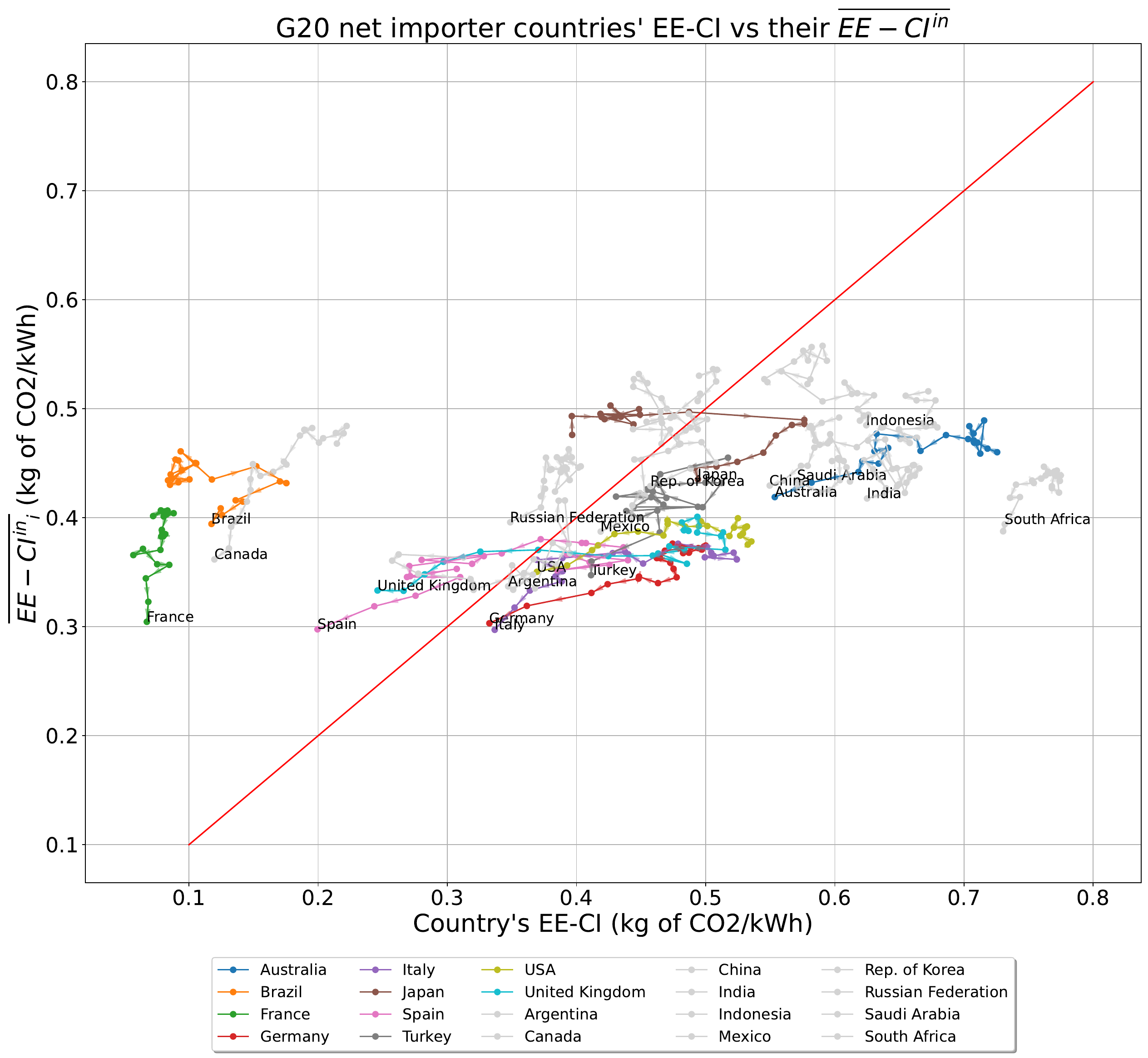}
\caption{Top panels: scattering the weighted mean of the GDP-CIs of each country `exporting partners' versus its own GDP-CI reveals that many `net exporters' are less economically efficient than the countries they export to (left panel); analogously, scattering the weighted mean of the GDP-CIs of each country `importing partners' versus its own GDP-CI reveals that many `net importers' are more economically efficient than the countries they import from (right panel). Bottom panels: scattering the weighted mean of the EE-CIs of each country `exporting partners' versus its own EE-CI reveals that many `net exporters' have an EE-CI that is steadily larger than the one of the neighbours they export to (left panel); analogously, scattering the weighted mean of the EE-CIs of each country `importing partners' versus its own EE-CI reveals that many `net importers' still have an EE-CI which is larger than the EE-CI of the neighbours they import from (right panel): their trajectories, however, are evolving towards the left, an outcome suggesting that their EE-CI has decreased, over the past twenty-one years, at a higher rate than that of their partners.}
\label{figB5}
\end{figure*}

The top-left panel of fig. \ref{figB5} shows the evolution of the set of values $\left\{\overline{\text{GDP-CI}}^{out}\right\}$, scattered versus the set of values $\{\text{GDP-CI}\}$, for the `net exporters': as it can be appreciated, most of their GDP-CIs are steadily larger than the GDP-CI of the neighbours they export to. The top-right panel of fig. \ref{figB5} shows the evolution of the set of values $\left\{\overline{\text{GDP-CI}}^{in}\right\}$, scattered versus the set of values $\{\text{GDP-CI}\}$, for the `net importers': in this case, most of their GDP-CIs are steadily smaller than the GDP-CI of the neighbours they import from.

The bottom panels of fig. \ref{figB5} show the evolution of the set of values $\left\{\overline{\text{EE-CI}}^{out}\right\}$, scattered versus the set of values $\{\text{EE-CI}\}$ and the evolution of the set of values $\left\{\overline{\text{EE-CI}}^{in}\right\}$, scattered versus the set of values $\{\text{EE-CI}\}$ for the same sets of countries as above. Many of the trajectories of `net exporters' lie below the identity line; although many of the trajectories of `net importers' lie below the identity line as well, it should be noticed that they evolve in a horizontal fashion, moving from right to left, an outcome suggesting that their EE-CI has decreased over the past twenty-one years while that of their partners is not. Overall, this leads us to conclude that countries whose export exceeds the import, export towards `cleaner' countries - equivalently, countries whose import exceeds the export, import from `less clean' countries.

\end{document}